# Timely Common Knowledge

## Characterising Asymmetric Distributed Coordination via Vectorial Fixed Points

Yannai A. Gonczarowski
Einstein Institute of Mathematics and
Center for the Study of Rationality
Hebrew University of Jerusalem
Jerusalem 91904, Israel
yannai@gonch.name

Yoram Moses
Department of Electrical Engineering
Technion—Israel Institute of Technology
Haifa 32000, Israel
moses@ee.technion.ac.il

## ABSTRACT

Coordinating activities at different sites of a multi-agent system typically imposes epistemic constraints on the participants. Specifying explicit bounds on the relative times at which actions are performed induces combined temporal and epistemic constraints on when agents can perform their actions. This paper characterises the interactive epistemic state that arises when actions must meet particular temporal constraints. The new state, called *timely common knowledge*, generalizes common knowledge, as well as other variants of common knowledge. While known variants of common knowledge are defined in terms of a fixed point of an epistemic formula, timely common knowledge is defined in terms of a *vectorial* fixed point of temporal-epistemic formulae. A general class of coordination tasks with timing constraints is defined, and timely common knowledge is used to characterise both solvability and optimal solutions of such tasks. Moreover, it is shown that under natural conditions, timely common knowledge is equivalent to an infinite conjunction of temporal-epistemic formulae, in analogy to the popular definition of common knowledge.

## Categories and Subject Descriptors

[**Artificial intelligence**]: Knowledge representation and reasoning — *Reasoning about belief and knowledge, Temporal reasoning, Causal reasoning and diagnostics*; [**Artificial intelligence**]: Distributed artificial intelligence — *Cooperation and coordination, multi-agent systems*; [**Real-time systems**]: Real-time system specification; [**Distributed computing methodologies**]

## 1. INTRODUCTION

The fact that knowledge is closely related to coordinated action in distributed and multi-agent systems is well established by now. Ensuring that actions are performed in linear temporal order requires the agents to obtain appropriate nested knowledge (knowledge about knowledge) [5], while coordinating simultaneous actions requires attaining common knowledge of particular facts [17]. The latter connection has found uses in the analysis of distributed protocols (see, e.g. [17, 11, 28]). One of the contributions of [17] was in relating approximations of simultaneous coordination to weaker variants of common knowledge, called *epsilon*-common knowledge and *eventual* common knowledge. While common knowledge is typically defined and thought of as an infinite conjunction of nested knowledge formulae, it may also be defined as a fixed point [3, 8]. The variants of common knowledge defined by Halpern and Moses in [17] are most naturally obtained by appropriately modifying the fixed-point definition of common knowledge. All of the forms of coordination analyzed in [17] are symmetric in nature, in the sense that they are invariant under renaming of agents. For example, $\varepsilon$-common knowledge arises when the agents are guaranteed to act at most $\varepsilon$ time units apart. In many natural situations, however, asymmetric forms of coordination arise. Let us consider an example.

EXAMPLE 1.1 (ROBOTIC CAR WASH).
*In an automated robotic car-wash enterprise, there are two washing robots $L$ and $R$, (with $L$ fitted to soap & rinse the left sides of cars, and $R$ fitted to soap & rinse the right sides), and one drying robot, denoted $D$. At some point after a car enters, it must be soaped & rinsed from both sides, and then dried. The robot $L$ is a new model, which takes only 4 minutes to perform its duty, while $R$ is an older model, requiring 6 minutes. The drying is applied to the whole car, and it must commence only after washing of both sides is complete. Moreover, drying should not begin more than 5 minutes after the first of the washing robots finishes rinsing the car, as water stains might otherwise incur. It follows that, in particular, no more than 5 minutes may elapse between the time at which the rinsing of the car's left side ends and the time at which the rinsing of its right side ends. This, in turn, implies that $L$ must start washing the car no later than 7 minutes after — and no more than 3 minutes before — $R$ starts washing it. Finally, it is obviously desirable to minimize the time that the car spends in the Car Wash.*

The temporal constraints in the car wash example make the design of the robots' control (the protocol that they follow) a delicate matter. With respect to a given car, each of the robots has only one decision to make: when to start treating the car — we shall refer to this as *the robot's action*. The times at which the robots act must satisfy the interactive constraints implied by the example. Clearly, the decision to act depends on when each of the other robots can (and will) commence treatment of this car. Before it can act, a robot must know (i.e., be sure) that the others will act in time, which requires, in particular, that the others will in turn know that *they* can act. More concretely, in our example, when $L$ starts washing a car, it must know that between 7 minutes earlier and 3 minutes later, $R$ will have started washing it, and that between 4 and $4 + 5 = 9$ minutes afterward, the drying robot $D$ will have started drying it. Conversely yet asymmetrically, when $R$ starts washing a car, it must know that between 3 minutes earlier and 7 minutes later, $L$ will have started washing it and that between 6 and $6 + 5 = 11$ minutes afterward, $D$ will have started drying it. We can similarly calculate $D$'s required knowledge about $L$ and $R$. Notice that this dependence is asymmetric — each robot calculates different bounds between its action and those of the two others.

The above discussion suggests that the robots in our example must reach some form of "temporal-epistemic equilibrium" in order to act. More generally, analogous situations seem to arise whenever a set of agents must coordinate their actions in a manner satisfying possibly asymmetric timing constraints. Our purpose in this paper is to concisely and usefully capture this form of interdependence in coordinated action. We shall do this by defining a new epistemic condition called *timely* common knowledge, which is, in a precise sense, necessary and sufficient for coordination as in the above example. Timely common knowledge generalizes and significantly extends common knowledge and its popular variants. Mathematically, the new notion is formally captured by way of a *vectorial* fixed point. Whereas common knowledge of an event $\psi$ can be defined as the greatest fixed point of the function $x \mapsto E(\psi \wedge x)$, mapping events to events, where $E$ is the operator denoting "everyone knows that...", a vectorial fixed point is the fixed point of a function mapping tuples of events to tuples of events. To our knowledge, such a technique has never before been utilized with regard to epistemic analysis. Roughly speaking, in the case of the car wash example, let $\bar{\xi} = (\xi_l, \xi_r, \xi_d)$ be the greatest fixed point of the function

$$\begin{pmatrix} x_l \\ x_r \\ x_d \end{pmatrix} \mapsto \begin{pmatrix} K_l(\ \psi_c\ \wedge\ \odot^{\leq 3} x_r\ \wedge\ \odot^{\leq 9} x_d\ ) \\ K_r(\ \psi_c\ \wedge\ \odot^{\leq 7} x_l\ \wedge\ \odot^{\leq 11} x_d\ ) \\ K_d(\ \psi_c\ \wedge\ \odot^{\leq -4} x_l\ \wedge\ \odot^{\leq -6} x_r\ ) \end{pmatrix},$$

where $\psi_c$ is the event "the car is here", where $K_i$ denotes "$i$ knows that..." and where $\odot^{\leq \varepsilon} x$ means that "$x$ holds at some (past, present or future) point in time, no later than $\varepsilon$ minutes after the current time". In the fixed point $\bar{\xi}$, robot $L$'s coordinate $\xi_l$ holds iff $K_l(\psi_c \wedge \odot^{\leq 3} \xi_r \wedge \odot^{\leq 9} \xi_d)$ does, and similarly for the other coordinates. Our results imply that the car-wash problem may be solved by having each robot $i$ perform its task as soon as $\xi_i$ holds, and that this solution is, in a precise sense, time-optimal. Roughly speaking, the tuple of events $\bar{\xi}$ will constitute *timely common knowledge* of $\psi_c$ (with respect to the timing constraints of Example 1.1).[1] Notice that $\bar{\xi}$ does not correspond to a single fact (or event) that may be true or false at a single point in time. Rather, it represents a tuple of facts, one for each agent of interest. Each of the facts should hold at its own individual time, and the different times jointly satisfy the conditions in the fixed point definition.

---

[1] The definition of timely common knowledge is made in Section 4 with respect to general timing constraints, and is, naturally, more subtle.

In Section 4, we relate timely common knowledge to coordination. We define a class of *timely coordination* specifications, in which actions by various agents must satisfy timing conditions as in the Car Wash example. Timely coordination allows both symmetric and asymmetric forms of communication, and it strictly generalizes many symmetric forms of coordination previously studied in the literature. We also show, in a precise sense, that timely common knowledge strictly generalizes standard common knowledge and some of its variants. In Section 6, we show another close connection between timely common knowledge and standard common knowledge. Recall that common knowledge is often described as an infinite conjunction of nested knowledge formulae. A temporal-epistemic variant applies in the case of timely common knowledge. Roughly speaking, consider the point $p$ at which $L$ acts in the above car wash example. Recall that $\psi_c$ denotes the fact that the car $c$ has arrived, then clearly $K_l \psi_c$ must hold at $p$. It is not hard to check that $K_l \odot^{\leq 3} K_r \psi_c$ should also hold at $p$, as should $K_l \odot^{\leq 3} K_r \odot^{\leq 7} K_l \psi_c$. Indeed, it is possible to generate arbitrarily deeply nested formulae that must hold at $p$. A different set of formulae must hold when $R$ acts, and yet another set when $D$ does. Thus, timely common knowledge implies an infinite set of nested formulae at each point of action. We show that it is in fact equivalent to a tuple of such sets under natural assumptions.

As an example of a natural application of our analysis, in Section 5 we present and mathematically analyze *timely-coordinated response* — a novel class of multi-agent coordination tasks. Roughly speaking, a timely-coordinated response task involves a prespecified *triggering* event $\psi$, such as the activation of a smoke detector or the arrival of a car to the car-wash facility. Should the trigger $\psi$ occur, then each agent $i$ in a set $I$ of agents should perform an action (called its *response* to $\psi$) specified by the task, and the timing of the actions must satisfy a constraint of the following form: for all $i, j \in I$, if $i$ acts at time $t_i$ and $j$ at $t_j$, then $-\delta(j, i) \leq t_j - t_i \leq \delta(i, j)$. The trigger $\psi$, the response actions, and the bounds $\delta$ are parameters specified in a given task. E.g. in the car wash example, the trigger is a car's arrival $\psi_c$, responses are robots' initiating their respective services, while $\delta(L, R) = 3$, $\delta(L, D) = 9$, $\delta(R, L) = 7$, $\delta(R, D) = 11$, $\delta(D, L) = -4$ and $\delta(D, R) = -6$. Timely-coordinated response is inspired by, and strictly generalizes, the response problems presented and studied by Ben-Zvi and Moses [5, 4, 6, 7].

We show that timely common knowledge is, in a precise sense, the epistemic counterpart of timely coordination. We use timely common knowledge to phrase a necessary and sufficient condition characterising protocols solving timely-coordinated response. Moreover, we show how timely common knowledge can be used to give a general technique for deriving a time-optimal solution (i.e. an optimal protocol) for any instance of timely-coordinated response.

The main contributions of this paper are:

- The theory of coordination in multi-agent systems is extended to treat timely coordination, in which general interdependent constraints are allowed;

- Timely common knowledge is defined as a vectorial fixed point and the mathematical soundness and key

- properties of its definition are established;

- The solvability of, and optimal solutions to, a general class of timely coordination tasks are characterised using timely common knowledge;

- Timely common knowledge is shown to generalize common knowledge and many of its variants; and

- Timely common knowledge is shown to be equivalent to an infinite conjunction under natural assumptions.

## 2. RELATED WORK

The notion of common knowledge was defined by the philosopher David Lewis in [24]. Its relevance to game theory was shown by Aumann [2] and to AI by McCarthy [26]. Halpern and Moses [17] introduced it to distributed computing, showed its connection to simultaneity, and defined weaker variants of common knowledge corresponding to "approximations" of simultaneity. Common knowledge and its variants have had various applications in distributed computing [9, 11, 28, 19, 22, 12]. More recently, Ben-Zvi and Moses studied how time bounds on message transmission impact coordination in message-passing systems [5, 6, 4]. Most of their work studied coordination problems that are specified by partial orders. In [7], Ben-Zvi and Moses consider a notion of *tightly-timed* coordination in which agents act at precise time differences from each other. This gives rise to a generalization of common knowledge in which agents are considered at different prespecified times. All fixed-point epistemic notions (common knowledge and its variants) in the above works are based on a standard (scalar) fixed-point definition. The analysis in this paper significantly extends the connection between coordination and epistemic notions.

## 3. MODEL AND NOTATION

For ease of exposition, we adopt the multi-agent systems model, based on contexts, runs and systems, of Fagin *et al.* [12]. The model captures the possible histories, called runs, of a finite set of agents $\mathbb{I}$. We model time as being discrete, ranging over the set $\mathbb{T} = \mathbb{N} \cup \{0\}$ of nonnegative integers.[2] Each agent $i \in \mathbb{I}$ may be thought of as an automaton, existing at any specific time $t \in \mathbb{T}$ in one of a set of possible states $\mathbb{L}_i$. The set of possible global states of the model, describing a snapshot of the system at some given time, is thus $\mathbb{L}_e \times \bigtimes_{i \in \mathbb{I}} \mathbb{L}_i$, where $\mathbb{L}_e$ is a set of possible states for the environment. We denote by $\mathcal{R}$ the set of possible runs, or possible histories, of the model, where a run $r \in \mathcal{R}$ is a function $r : \mathbb{T} \to \mathbb{L}_e \times \bigtimes_{i \in \mathbb{I}} \mathbb{L}_i$, from times to global states. A *point* is a run-time pair $p = (r, t) \in \mathcal{R} \times \mathbb{T}$, denoting time $t$ in the run $r$. The local state of an agent $i \in \mathbb{I}$ at the point $(r, t)$ is denoted by $r_i(t)$. We denote by $\mathbb{P}$ the set of protocols, where a protocol $P \in \mathbb{P}$ is a tuple $P = (P_i)_{i \in \mathbb{I}}$, in which each $P_i$ is a function from the set $\mathbb{L}_i$ of $i$'s local states to sets of possible actions (or to a single option, if $P$ is deterministic) for the actions to be performed by $i$ when at that state. Finally, a context $\gamma$ is a specification of a protocol for the environment, possible initial global states, any relevant constraints on runs (e.g. an agent may not perform two certain given actions at the same time), and a transition function from the global state and all actions performed at any time $t$, to the global state at $t+1$. For a context $\gamma$ and a protocol $P \in \mathbb{P}$, we denote by $R(P, \gamma)$ the set of runs of $P$ in $\gamma$.

### 3.1 Events, Knowledge Operators and Temporal Operators

There are two equivalent ways of defining knowledge in systems, one in terms of propositions and modal operators in modal logic [12], and the other, proposed by Aumann [2], in terms of events and of functions on events. We follow the latter, since it facilitates the formulation of fixed points, which play a role in our analysis. Informally, however, we use the terms *fact* and *event* interchangeably. As in probability theory, we represent events using the set of points at which they hold. A set of runs $R$ gives rise to a *(R-)universe* $\Omega_R \triangleq R \times \mathbb{T}$, and a corresponding $\sigma$-algebra of events $\mathcal{F}_R \triangleq 2^{\Omega_R}$. Thus, for example, the event "agent $i$ is performing action $\alpha$", is formally associated with all points $(r, t) \in \Omega_R$ at which $i$ performs $\alpha$.

We make use of several temporal operators applied to events. These are very much in the spirit of standard linear-time operators (see Manna and Pnueli [25]), except that in our case two of the operators may refer to the past as well as the future. We thus use slight variations on the standard symbols. A few basic properties of these operators are explored in Appendix C. We define three temporal operators as functions $\mathcal{F}_R \to \mathcal{F}_R$ as follows;[3] fix $R \subseteq \mathcal{R}$ and let $\psi \in \mathcal{F}_R$.

- $\diamondsuit \psi \triangleq \{(r, t) \in \Omega_R \mid \exists t' \in \mathbb{T} : (r, t') \in \psi\}$; the event "$\psi$ holds at some past, present or future time (during the current run)",

- $\odot^{\varepsilon} \psi \triangleq \{(r, t) \in \Omega_R \mid (r, t + \varepsilon) \in \psi\}$, for $\varepsilon \in \mathbb{Z}$; the event "$\psi$ holds at *exactly* $\varepsilon$ time units from now", and

- $\odot^{\leq \varepsilon} \psi \triangleq \left\{(r, t) \in \Omega_R \,\middle|\, \exists t' \in \mathbb{T} : \begin{array}{l} t' \leq t + \varepsilon \ \& \\ (r, t') \in \psi \end{array}\right\}$, for $\varepsilon \in \mathbb{Z} \cup \{\infty\}$; the event "$\psi$ holds at some (past, present, or future) time, no later than $\varepsilon$ time units from now".

The standard definition of knowledge in this setting is also a function on events. Intuitively, an agent's information is captured by its local state $r_i(t)$. Accordingly, two points $(r, t)$ and $(r', t')$ are considered *indistinguishable* in the eyes of $i$ if $i$'s local state at both points is the same. We use $K_i$ to denote $i$'s knowledge, and define the event "$i$ knows $\psi$" by

- $K_i \psi \triangleq \{(r, t) \in \Omega_R \mid (r', t') \in \psi \text{ whenever } r'_i(t) = r_i(t)\}$.

Since $K_i \psi$ is itself an event, nested knowledge facts such as $K_j K_i \psi$ are immediately well defined. This gives rise to a standard S5 notion of knowledge, equivalent to the standard definition in terms of partitions. See Appendix A for a discussion, and for a definition of common knowledge.

---

[2] All results in this paper hold verbatim if we consider infinite sets $\mathbb{I}$ of agents, and with only trivial changes if time is continuous, so that $\mathbb{T} \triangleq \mathbb{R}_{\geq 0}$. We avoid modifying the model to handle continuous time for ease of exposition.

[3] While the following definitions depend on $R$, we omit $R$ from these notations for readability, as the set of runs will be clear from the discussion. We follow this convention when presenting some other definitions below as well.

## 3.2 Event Ensembles

Roughly speaking, it is possible for an agent $i$ to act precisely whenever an event $\psi \in \Omega_R$ occurs, only if at every point at which $\psi$ holds, $i$ knows that it does, i.e. if $\psi = K_i \psi$. Such an event is said to be *i-local*. Equivalently, $\psi$ is *i-local* if its truth is determined by $i$'s local state, i.e. if there exists $S \subseteq \mathbb{L}_i$ s.t. for every $(r,t) \in \Omega_R$, we have $(r,t) \in \psi$ iff $r_i(t) \in S$. In the study of coordination, we are usually interested in the interaction between the actions of several agents. Consider, for example, a scenario in which two agents, Alice and Bob, must perform two respective actions $\alpha$ and $\beta$ in some coordinated manner. Then the set $\boldsymbol{e}_A$ of points at which Alice performs $\alpha$ is a local event for Alice, and likewise for the corresponding set $\boldsymbol{e}_B$ for Bob and $\beta$. The pair $\bar{\boldsymbol{e}} \triangleq (\boldsymbol{e}_A, \boldsymbol{e}_B)$ is called an *ensemble* for Alice and Bob. More generally, following Fagin *et al.*, given a set of agents $I \subseteq \mathbb{I}$, we define an *I-ensemble* to be an *I*-tuple of events $\bar{\boldsymbol{e}} = (\boldsymbol{e}_i)_{i \in I} \in \mathcal{F}_R{}^I$, in which $\boldsymbol{e}_i$ is *i*-local, for each $i \in I$. Returning to Alice and Bob, consider a deterministic protocol in which whenever Alice performs action $\alpha$, Bob is guaranteed to simultaneously perform action $\beta$ and vice versa. Since $\alpha$ and $\beta$ are guaranteed to be simultaneous actions, we have $\boldsymbol{e}_A = \boldsymbol{e}_B$. An ensemble $\bar{\boldsymbol{e}}$ with this property is thus said to be *perfectly coordinated*. Fagin *et al.* [13] have studied the properties of such ensembles, as well as of ensembles satisfying weaker forms of coordination (eventual coordination and $\varepsilon$-coordination) defined in [17]. See Appendix B.1 for more details.

## 4. TIMELY COORDINATION & TIMELY COMMON KNOWLEDGE

Given a set of agents $I$, we denote by the set of distinct pairs of agents in $I$ by $I^{\tilde{2}} \triangleq \{(i,j) \in I^2 \mid i \neq j\}$. We define a *timely-coordination spec* to be a pair $(I, \delta)$, where $I \subseteq \mathbb{I}$ is a set of agents and $\delta : I^{\tilde{2}} \to \mathbb{Z} \cup \{\infty\}$. Intuitively, $\delta(i,j)$ denotes an upper bound on the time from when $i$ performs her action, until when $j$ performs his.[4] We can now formally define timely coordination.

DEFINITION 4.1 (TIMELY-COORDINATION).
*Given a timely-coordination spec $(I,\delta)$ and a system $R \subseteq \mathcal{R}$, we say that an I-ensemble $\bar{\boldsymbol{e}} \in \mathcal{F}_R{}^I$ is $\boldsymbol{\delta}$-coordinated (in R) if for every $(i,j) \in I^{\tilde{2}}$ and for every $(r,t) \in \boldsymbol{e}_i$, there exists $t' \le t + \delta(i,j)$ s.t. $(r,t') \in \boldsymbol{e}_j$.*

While, as discussed in Appendix A, the popular definition of common knowledge is in terms of an infinite conjunction of nested knowledge formulae, Barwise [3], following Harman [8], has defined common knowledge as a fixed point. Indeed, if we denote $E_I \psi = \bigcap_{i \in I} K_i \psi$ ("everybody in $I$ knows"), then the following is an equivalent way of formulating common knowledge as a fixed point.[5]

THEOREM 4.2 ([17]). *Let $R \subseteq \mathcal{R}$ and $I \subseteq \mathbb{I}$. Then $C_I \psi$ is the greatest fixed point of the function $f_\psi : \mathcal{F}_R \to \mathcal{F}_R$ given by $x \mapsto E_I(\psi \cap x)$, for every event $\psi \in \mathcal{F}_R$.*

As mentioned in the introduction, a classic result [17], which stems from Theorem 4.2, is that common knowledge tightly relates to perfect coordination. One manifestation of this is in the fact that if an action $\alpha$ is guaranteed to be performed simultaneously by a set of agents whenever any of them performs it, then these agents must have common knowledge of the occurrence of $\alpha$ when it is performed. (Intuitively, the guaranteed simultaneity of $\alpha$ causes its joint occurrence to be inferred at once by all participants who perform it.) Conversely, whenever common knowledge of a fact arises among a set of agents, it does so simultaneously for all agents. See Appendix B.2 for further details, as well as a review of the analogous analysis for the weaker variants of common knowledge defined in [17]. Our purpose is to similarly relate timely coordination to an epistemic notion. Consider the points at which the robots act in the Car Wash example. In general, the robots may act at different times. Moreover, while the local events that the various robots must respectively know in order for them to act are interdependent, they differ from one another. Therefore, instead of seeking a fixed point of a function on (single events in) $\mathcal{F}_R$ as done for common knowledge and previous variants, we define a function on $\mathcal{F}_R{}^I$ — the set of $I$-tuples of events. Given an event $\psi \in \mathcal{F}_R$ and a timely-coordination spec $(I,\delta)$, we define a function $f_\psi^\delta$ on $\mathcal{F}_R{}^I$ in which each coordinate $i$ captures the respective constraints of the agent $i$, based on $\psi$ and $\delta$. The greatest fixed point of $f_\psi^\delta$, denoted by $C_I^\delta \psi$ (this is an $I$-tuple of events), is shown to capture timely coordination, and is thus the desired ensemble. Since $f_\psi^\delta$ is a function of several variables, it is a *vectorial* function, and its fixed point is a vectorial fixed point [1].[6]

### 4.1 Timely Common Knowledge as a Vectorial Fixed Point

We start by defining a lattice structure on $\mathcal{F}_R{}^I$. A *greatest fixed point* of a function $f$ on $\mathcal{F}_R{}^I$ is a fixed point of $f$ that is greater than any other fixed point thereof, according to the partial order $\le$ of the lattice. Recall that a member of $\mathcal{F}_R{}^I$ is a tuple of events of the form $\bar{\varphi} \triangleq (\varphi_i)_{i \in I}$.

DEFINITION 4.3 (LATTICE STRUCTURE ON $\mathcal{F}_R{}^I$).
*Let $R \subseteq \mathcal{R}$ and let $I \subseteq \mathbb{I}$. The following partial order relation and binary operations define a lattice structure on $\mathcal{F}_R{}^I$.*

- *Order:* $\bar{\varphi} \le \bar{\xi}$ *iff* $\forall i \in I : \varphi_i \subseteq \xi_i$.
- *Join:* $\bar{\varphi} \vee \bar{\xi} \triangleq (\varphi_i \cup \xi_i)_{i \in I}$.
- *Meet:* $\bar{\varphi} \wedge \bar{\xi} \triangleq (\varphi_i \cap \xi_i)_{i \in I}$.

We are now ready to define timely common knowledge.

DEFINITION 4.4 (TIMELY COMMON KNOWLEDGE).
*Let $R \subseteq \mathcal{R}$ and let $(I,\delta)$ be a timely-coordination spec. For each $\psi \in \mathcal{F}_R$, we define $\boldsymbol{\delta}$-common knowledge of $\psi$ by $I$, denoted by $C_I^\delta \psi$, to be the greatest fixed point of the function $f_\psi^\delta : \mathcal{F}_R{}^I \to \mathcal{F}_R{}^I$ given by*

$$f_\psi^\delta : \quad (x_i)_{i \in I} \quad \mapsto \quad \left( K_i \Big( \psi \cap \bigcap_{j \in I \setminus \{i\}} \odot^{\le \delta(i,j)} x_j \Big) \right)_{i \in I}.$$

---

[4] If time were continuous (i.e. $\mathbb{T} = \mathbb{R}_{\ge 0}$), then the range of $\delta$ would be $(\mathbb{T} - \mathbb{T}) \cup \{\infty\} = \mathbb{R} \cup \{\infty\} = (-\infty, \infty]$.
[5] The equivalence is in the standard models; see Barwise [3] for a discussion of various accepted definitions for common knowledge and of models in which they do not coincide.
[6] While vectorial fixed points may alternatively be captured by nested fixed points [1, Chapter 1], in our case we argue that the vectorial representation better parallels the underlying intuition. We are not aware of either vectorial, or nested fixed points being used in an epistemic setting before.

We justify Definition 4.4 in three steps. First, we show that $C_I^\delta \psi$ is well-defined and satisfies a natural induction rule and a monotonicity property. (For proofs of all propositions given in this paper, see Appendix C.)

LEMMA 4.5. *Let $(I, \delta)$ be a timely-coordination spec, let $R \subseteq \mathcal{R}$ and let $\psi \in \mathcal{F}_R$.*

1. *$C_I^\delta \psi$ is well-defined, i.e. $f_\psi^\delta$ has a greatest fixed point.*

2. *Induction Rule: Every $\bar{\xi} \in \mathcal{F}_R{}^I$ satisfying $\bar{\xi} \leq f_\psi^\delta(\bar{\xi})$ also satisfies $\bar{\xi} \leq C_I^\delta \psi$.*

3. *$C_I^\delta$ is monotone: $\psi \subseteq \phi \Rightarrow C_I^\delta \psi \leq C_I^\delta \phi$, for every $\psi, \phi \in \mathcal{F}_R$.*

The induction rule is a powerful tool for analyzing situations giving rise to timely common knowledge. It states that if $\xi_i$ implies the $K_i$ statement in Definition 4.4, with $x_j$ substituted by $\xi_j$ everywhere, then each agent $i$ knows its respective coordinate of $C_I^\delta \psi$ whenever $\xi_i$ holds.

A timely-coordination spec is a fairly general tool for defining relative timing constraints. Particular simple instances can capture previously studied forms of coordination. Namely, if $\delta \equiv \infty$, timely coordination coincides with eventual coordination, and for any $\varepsilon < \infty$, the form of coordination obtained by setting $\delta \equiv \varepsilon$ closely relates to $\varepsilon$-coordination (and hence to perfect coordination when $\delta \equiv 0$). Indeed, for coordinate-wise stable ensembles (see Appendix C.4) and for ensembles with at most a single point per agent per run (see Section 5 for an example), $\delta \equiv \varepsilon$ precisely captures $\varepsilon$-coordination and $\delta \equiv 0$ specifies perfect coordination. Furthermore, timely common knowledge is closely related to the corresponding variants of common knowledge, for each of these special cases of constant $\delta$. (See Appendix D.2 for the precise details.) Our second step is to show that timely common knowledge closely corresponds to timely coordination, in the same sense in which common knowledge corresponds to perfect coordination, and variants of common knowledge to their respective forms of coordination. (See, once again, Appendix B.2.) The following theorem establishes this correspondence. (While phrasing this theorem, and henceforth, we use the shorthand notation $\cup \bar{\xi} \triangleq \bigcup_{i \in I} \xi_i$, for every $\bar{\xi} = (\xi_i)_{i \in I} \in \mathcal{F}_R{}^I$.)

THEOREM 4.6. *Let $R \subseteq \mathcal{R}$ and let $(I, \delta)$ be a timely-coordination spec.*

1. *$C_I^\delta \psi$ constitutes a $\delta$-coordinated $I$-ensemble, for every $\psi \in \mathcal{F}_R$.*

2. *$\cup C_I^\delta \psi \subseteq \psi$, for every $\psi \in \mathcal{F}_R$.*

3. *If $\bar{e} \in \mathcal{F}_R{}^I$ is a $\delta$-coordinated $I$-ensemble satisfying $\cup \bar{e} \subseteq \psi$ for some $\psi \in \mathcal{F}_R$, then $\bar{e} \leq C_I^\delta \psi$.*

4. *If $\bar{e} \in \mathcal{F}_R{}^I$ is a $\delta$-coordinated $I$-ensemble, then $\bar{e} \leq C_I^\delta(\cup \bar{e})$.*

5. *If $\bar{e} \in \mathcal{F}_R{}^I$ is a $\delta$-coordinated $I$-ensemble, then $\cup \bar{e} = \cup C_I^\delta(\cup \bar{e})$.*

Theorem 4.6 highlights some key properties of the fundamental connection between $\delta$-coordination and $\delta$-common knowledge: (Parts 1, 4 and 5 are analogues of Theorems B.4, B.5 and B.6, the latter part being stronger in a sense than its counterparts from Theorems B.5 and B.6 regarding eventual- and $\varepsilon$-common knowledge, respectively.) Parts 1–3 characterise $\delta$-common knowledge of $\psi$ as the greatest $\delta$-coordinated event ensemble that implies $\psi$.[7] Moreover, Part 3 provides convenient means to prove that timely common knowledge holds. Part 4 says that regardless of the way a $\delta$-coordinated ensemble is formed (be it using $\delta$-common knowledge of some event $\psi$, or otherwise), the fact that its $i$'th coordinate holds implies that the $i$'th coordinate of $\delta$-common knowledge of (the disjunction of) this ensemble holds as well. Finally, part 5 captures the fact that the union of any $\delta$-coordinated ensemble is a fixed point of $\cup C_I^\delta$, and, together with Part 1, implies the idempotence of $\cup C_I^\delta$. Our third step is demonstrating the usefulness of timely common knowledge, which we do in the next section.

## 5. TIMELY-COORDINATED RESPONSE

We now harness the machinery developed in the previous sections to study a class of coordination problems. In these problems, the occurrence of a particular event $\phi$ must trigger responses by a set $I \subseteq \mathbb{I}$ of agents, and the responses must be timely coordinated according to a given spec $\delta$.[8] The triggering event $\phi$ may be the arrival of a car at the Car Wash, the ringing of a smoke alarm, or some other event that requires a response. A run $r$ during which $\phi$ occurs (i.e. $(r, t) \in \phi$ for some $t \in \mathbb{T}$) is called $\phi$-*triggered*. Following in the spirit of [5] and generalizing their definitions (see Appendix D.1), we define this class of coordination problems as follows.

DEFINITION 5.1 (TIMELY-COORDINATED RESPONSE).
*A **timely-coordinated response** problem, or TCR, is a quintuplet $\tau = (\gamma, \phi, I, \delta, \bar{\alpha})$, where $\gamma$ is a context, $\phi \in \mathcal{F}_\mathcal{R}$ is an event, $(I, \delta)$ is a timely-coordination spec and $\bar{\alpha} = (\alpha_i)_{i \in I}$ is a tuple of actions, one for each $i \in I$. A protocol $P$ is said to **solve** a TCR $\tau = (\gamma, \phi, I, \delta, \bar{\alpha})$ if for every $r \in R(P, \gamma)$,*

- *If $r$ is $\phi$-triggered and $\phi$ first occurs in $r$ at $t_\phi \in \mathbb{T}$, then each agent $i \in I$ responds (i.e. performs $\alpha_i$) in $r$ exactly once, at a time $t_i \geq t_\phi$ s.t. for every $(i, j) \in I^2$, it holds that $t_j \leq t_i + \delta(i, j)$.*

- *If $r$ is not $\phi$-triggered, then none of the agents in $I$ respond in $r$.*

We say that $\tau$ is *solvable* if there exists a protocol $P \in \mathbb{P}$ that solves it. We now show that attaining timely common knowledge is a necessary condition for action in a protocol solving timely-coordinated response, in the sense that an agent cannot respond unless is has attained its respective coordinate of timely common knowledge.[9] Indeed, Theo-

---

[7] Neither eventual- nor $\varepsilon$-common knowledge give way for a clean analogous characterisation. (See Appendix D.2 for more details.)

[8] For ease of exposition, we assume that each agent is associated with exactly one action. Essentially the same analysis applies if we allow each agent to be associated with more than one response action.

[9] In the following propositions, we work in the universe $\Omega_{R(P,\gamma)}$ defined by the system of runs of the given protocol in question. All knowledge and temporal operators are therefore relative to this universe. Furthermore, we slightly abuse notation by writing $\phi$ to refer to $\phi \cap \Omega_{R(P,\gamma)}$, i.e. the restriction of $\phi$ to this universe.

rem 4.6(3) implies:[10]

COROLLARY 5.2. *Let $P \in \mathbb{P}$ be a protocol solving a TCR $\tau = (\gamma, \phi, I, \delta, \bar{\alpha})$, and let $r \in R(P, \gamma)$ be a $\phi$-triggered run. If $i \in I$ responds at time $t_i$ in $r$, then $(r, t_i) \in \left(C_I^\delta(\odot^{\leq 0}\phi)\right)_i$.*

In fact, timely common knowledge is not only necessary for solving a TCR, but also *sufficient* for doing so. (See below.) Indeed, we now argue that timely common knowledge can be used to design time-optimal solutions for arbitrary TCRs. For the notion of time-optimality to be well defined, we define it with regard to each family of protocols that are the same in all aspects, except for possibly the time at which (and whether) agents respond. To this end, we restrict ourselves to protocols that may be represented as a pair $P = (P_{-\bar{\alpha}}, P_{\bar{\alpha}})$, s.t. the output of $P$ is a Cartesian product of the outputs of its two parts, where $P_{\bar{\alpha}}$ specifies whether to respond, while $P_{-\bar{\alpha}}$ specifies everything else. (Natural examples for such protocols are those in which the choice of whether to respond is deterministic.) Furthermore, we restrict ourselves to contexts in which none of $\bar{\alpha}$ affect the agents' transitions in any way (and hence do not affect any future states or actions). Under these conditions, given two protocols $P = (P_{-\bar{\alpha}}, P_{\bar{\alpha}})$ and $P' = (P_{-\bar{\alpha}}, P'_{\bar{\alpha}})$ that share same non-response component $P_{-\bar{\alpha}}$, there exists a natural isomorphism $\sigma : R(P, \gamma) \xrightarrow{\sim} R(P', \gamma)$, in which corresponding runs agree in all aspects except for possibly the times at which (and whether) responses are performed; we thus say that two such protocols are *run-equivalent*. Furthermore, we slightly abuse notation by writing $R(P_{-\bar{\alpha}}, \gamma)$ to refer to both $R(P, \gamma)$ and $R(P', \gamma)$, which coincide using $\sigma$. We say that a protocol $P = (P_{-\bar{\alpha}}, P_{\bar{\alpha}})$ is a *time-optimal* solution for a TCR $\tau$ if $P$ solves $\tau$ and, moreover, responses are never performed in $P$ later than in any solution $P'$ of $\tau$ that is run-equivalent to $P$. More formally, we demand that for every $\phi$-triggered $r \in R(P, \gamma)$ and for every $i \in I$, if $i$ responds at time $t_i$ in $r$ and at time $t'_i$ in $\sigma(r) \in R(P', \gamma)$ (with $\sigma$ as above), then necessarily $t_i \leq t'_i$. It should be noted that it is not *a priori* clear that TCRs admit time-optimal solutions. We now show not only that all solvable TCRs do, but moreover, that for every solution there exists a run-equivalent time-optimal solution and that all time-optimal solutions have each agent responding at the first instant at which it attains its respective coordinate of timely common knowledge.[11]

COROLLARY 5.3. *Let $\tau = (\gamma, \phi, I, \delta, \bar{\alpha})$ be a solvable TCR and let $P = (P_{-\bar{\alpha}}, P_{\bar{\alpha}})$ be a protocol solving it. The run-equivalent protocol $P' = (P_{-\bar{\alpha}}, P'_{\bar{\alpha}})$ in which every $i \in I$ responds at the first instant at which $\left(C_I^\delta(\odot^{\leq 0}\phi)\right)_i$ holds (in $\Omega_{R(P_{-\bar{\alpha}}, \gamma)}$), is a time-optimal solution for $\tau$.*

Indeed, we may now formalize our previous statement regarding timely common knowledge being necessary and sufficient for solving a TCR $\tau = (\gamma, \phi, I, \delta, \bar{\alpha})$: a protocol $P$ is run-equivalent to a solution of $\tau$ iff $C_I^\delta(\odot^{\leq 0}\phi)$ is attained in each of its $\phi$-triggered runs. (See Corollary C.13.)

---
[10] Observe that $\odot^{\leq 0}$ stands for the temporal operator "previously".
[11] As noted in Appendix C, in some runs of certain systems $R(P_{-\bar{\alpha}}, \gamma)$ in a continuous-time model, the set of times at which $\left(C_I^\delta(\odot^{\leq 0}\phi)\right)_i$ holds does not attain its infimum value. It is possible to similarly show that in such pathological cases, no time-optimal protocol that is run-equivalent to $P$ exists.

Attaining true (*not* timely) common knowledge of a fact of interest is often an effective and intuitive way of synchronization, which may also be used to solve TCRs. However, in addition to such a solution being suboptimal in many cases, timely common knowledge is often attainable even when common knowledge is not. In the Car Wash setting, for example, if the arrival of a car is guaranteed to be observed by each robot (privately) within at most 2 time units, then the TCR can be readily solved (and timely common knowledge attained) even though techniques of [17] may be used to show that the arrival of the car might never become common knowledge.

We conclude this section by noting that in contexts supporting full-information protocols (see, e.g. [12]), the above tools may be applied to obtain both a globally time-optimal solution to, as well as a solvability criterion for, arbitrary TCRs. We defer the details to the full paper.

## 6. A CONSTRUCTIVE DEFINITION FOR TIMELY COMMON KNOWLEDGE

The analysis of Section 5 provides us with time-optimal solutions for timely-coordinated response. The fly in the ointment, though, is how to implement these solutions, i.e. how to check whether a certain coordinate of timely common knowledge holds, given the state of the corresponding agent. We now take a step in this direction, which also sheds some more light on the fixed-point analysis of the previous section, and makes the notion of timely common knowledge more concrete. Under natural assumptions (see Theorem C.20 for details), we obtain, for every $i \in I$:

$$(C_I^\delta \psi)_i = \bigcap_{(i, i_2, \ldots, i_n) \in I^{\bar{*}}} K_i \odot^{\delta(i, i_2)} K_{i_2} \odot^{\delta(i_2, i_3)} K_{i_3} \cdots \odot^{\delta(i_{n-1}, i_n)} K_{i_n} \psi, \quad (1)$$

where $I^{\bar{*}} \triangleq \{(i_1, \ldots, i_n) \in I^* \mid \forall m : i_m \neq i_{m+1}\}$ denotes the set of all finite non-stuttering sequences of elements of $I$. Note that for $\delta \equiv 0$ (perfect coordination), (1) yields in each coordinate a familiar definition (see Observation A.4) of common knowledge as an infinite conjunction: (cf. the more popular Definition A.3, which is generalized by eventual- and $\varepsilon$-common knowledge, but is symmetric in nature, and therefore less natural for generalization in our setting.)

$$C_I \psi = \bigcap_{(i_1, \ldots, i_n) \in I^{\bar{*}}} K_{i_1} K_{i_2} K_{i_3} \cdots K_{i_n} \psi.$$

The formulation of timely common knowledge in terms of an infinite conjunction provides a constructive interpretation of the time-optimal solution from Corollary 5.3. Roughly speaking, each agent $i \in I$ should respond at the first instant at which all nested-knowledge formulae of the form $K_i \odot^{\delta(i, i_2)} K_{i_2} \odot^{\delta(i_2, i_3)} K_{i_3} \cdots \odot^{\delta(i_{n-1}, i_n)} K_{i_n} \odot^{\leq 0} \phi$ hold for all $(i, i_2, \ldots, i_n) \in I^{\bar{*}}$. (See Corollary C.22 for the precise phrasing.) While this may appear to take us a step closer to implementing time-optimal solutions, a naïve implementation may still require potentially infinitely many tests. In fact, as in the case of common knowledge, in practice timely common knowledge may be established using the induction rule of Theorem 4.6(3). We also refer the reader to [16, Chapters 6 and 9] for a study of the causal structure of these tests, which uses a different set of tools and which is, therefore, out of the scope of this paper.

## 7. CONCLUDING REMARKS

This paper suggests a broader connection between epistemic analysis and distributed coordination than was previously realized. The novel concept of timely common knowledge provides a formal connection between distributed protocols and a new form of equilibria, thus bringing distributed and multi-agent protocols closer to the realm of games, even in the absence of utilities and preferences. It should be noted, however, that the equilibrium in our analysis is not merely among strategies; in the Car Wash scenario, for example, the particular time instants at which the various robots act are at a *temporal-epistemic equilibrium*.

While this paper introduces vectorial fixed-point epistemic analysis as a tool for defining timely common knowledge, we believe that it will prove to be applicable well beyond the scope of problems considered here. We are currently pursuing generalizations and variations on the techniques presented in this paper for varying purposes, from generalizations of timely common knowledge to analyses of significantly different tasks, such as distributed agreement problems, which do not involve any form of timely coordination.

Fixed points, be they scalar or vectorial, be they temporal-epistemic or of any other kind, provide formal, yet intuitive, means of capturing equilibria in multi-agent systems. Many systems around us, from subatomic physical systems to astrophysical ones, and from animal societies to stock markets, exist in some form of equilibrium, possibly reached as a result of a long-forgotten spontaneous symmetry breaking. It is thus only natural to conjecture that fixed-point analyses of distributed algorithms and multi-agent systems hold the potential to provide significant further insights that are yet to be discovered.

## 8. ACKNOWLEDGMENTS

This work was supported in part by the Israel Science Foundation (ISF) under Grant 1520/11, and by the European Research Council under the European Community's Seventh Framework Programme (FP7/2007-2013) / ERC grant agreement no. [249159]. We would like to thank the reviewers for their useful comments. The first author would like to thank Gil Kalai, the co-advisor of his M.Sc. thesis [16]; this paper is based upon Chapters 7, 8 and 10 thereof.

# APPENDIX

## A. KNOWLEDGE AND COMMON KNOWLEDGE

We first survey a few immediate (and well-known) properties of the knowledge operator, which is defined in Section 3.

OBSERVATION A.1. *Let $R \subseteq \mathcal{R}$ and let $i \in \mathbb{I}$. By definition of $K_i$, we have:*

- *Knowledge Axiom: $K_i\psi \subseteq \psi$, for every $\psi \in \mathcal{F}_R$.*
- *Positive Introspection Axiom: $K_iK_i\psi = K_i\psi$, for every $\psi \in \mathcal{F}_R$.*
- *Monotonicity: $\psi \subseteq \phi \Rightarrow K_i\psi \subseteq K_i\phi$, for every $\psi, \phi \in \mathcal{F}_R$.*
- *$K_i$ commutes with intersection:*
  *$K_i(\cap\Psi) = \bigcap\{K_i\psi \mid \psi \in \Psi\}$, for every set of events $\Psi \subseteq \mathcal{F}_R$.*

We now build upon the definition of knowledge and define the notions of "everyone knows" and of "common knowledge".

DEFINITION A.2 (EVERYONE KNOWS). *Let $R \subseteq \mathcal{R}$ and let $I \subseteq \mathbb{I}$. For every $\psi \in \mathcal{F}_R$, denote $E_I\psi \triangleq \bigcap_{i \in I} K_i\psi$.*

One popular, constructive definition of common knowledge [14] is the following, defining that an event is common knowledge to a set of agents when all know it, all know that all know it, etc.

DEFINITION A.3 (COMMON KNOWLEDGE). *Let $R \subseteq \mathcal{R}$ and let $I \subseteq \mathbb{I}$. For every $\psi \in \mathcal{F}_R$, denote $C_I\psi \triangleq \bigcap_{n=1}^{\infty} E_I{}^n\psi$, where $E_I{}^0\psi \triangleq \psi$ and $E_I{}^n\psi \triangleq E_IE_I{}^{n-1}\psi$ for every $n \in \mathbb{N}$.*

OBSERVATION A.4. *Equivalently, by Definition A.2,*

$$C_I\psi = \bigcap_{(i_1,\ldots,i_n) \in I^*} K_{i_1}\cdots K_{i_n}\psi = \bigcap_{(i_1,\ldots,i_n) \in I^{\bar{*}}} K_{i_1}\cdots K_{i_n}\psi,$$

*where $I^{\bar{*}} \triangleq \{(i_1,\ldots,i_n) \in I^* \mid \forall m \in [n-1] : i_m \neq i_{m+1}\}$ denotes the set of all finite non-stuttering sequences of elements of $I$.*

## B. BACKGROUND: SYMMETRIC FORMS OF COORDINATION

In this section, we survey a few forms of coordination previously defined and analyzed by Halpern and Moses [17], as formulated for ensembles in [12, Section 11.6]. We reformulate these using events and adapt them to our notation.

### B.1 Definitions

DEFINITION B.1 (PERFECT COORDINATION). *Let $R \subseteq \mathcal{R}$ and let $I \subseteq \mathbb{I}$. An $I$-ensemble $\bar{e} \in \mathcal{F}_R{}^I$ is said to be **perfectly coordinated** (in $R$) if $e_i = e_j$ for every $i, j \in I$.*

DEFINITION B.2 (EVENTUAL COORDINATION [17, 12]). *Let $R \subseteq \mathcal{R}$ and let $I \subseteq \mathbb{I}$. An $I$-ensemble $\bar{e} \in \mathcal{F}_R{}^I$ is said to be **eventually coordinated** (in $R$) if for every $i, j \in I$ and for every $(r, t) \in e_i$, there exists $t' \in \mathbb{T}$ s.t. $(r, t') \in e_j$.*

DEFINITION B.3 ($\varepsilon$-COORDINATION [17, 12]). *Let $R \subseteq \mathcal{R}$, let $I \subseteq \mathbb{I}$ and let $\varepsilon \geq 0$. An $I$-ensemble $\bar{e} \in \mathcal{F}_R{}^I$ is said to be **$\varepsilon$-coordinated** (in $R$) if for every $i \in I$ and for every $(r, t) \in e_i$, there exists an interval $T \subseteq \mathbb{T}$ of length at most $\varepsilon$, s.t. $t \in T$ and s.t. for every $j \in I$ there exists $t' \in T$ s.t. $(r, t') \in e_j$.*

### B.2 Fixed-Point Analysis

While phrasing the propositions in this section, and henceforth, we use the shorthand notation $\cup\bar{\xi} \triangleq \bigcup_{i \in I} \xi_i$, for every $I$-ensemble $\bar{\xi} = (\xi_i)_{i \in I} \in \mathcal{F}_R{}^I$.

THEOREM B.4 ([17, 12]). *Let $R \subseteq \mathcal{R}$ and let $I \subseteq \mathbb{I}$.*

1. *$(C_I\psi)_{i \in I}$ constitutes a perfectly coordinated $I$-ensemble, for every $\psi \in \mathcal{F}_R$.*
2. *If $\bar{e} \in \mathcal{F}_R{}^I$ is a perfectly-coordinated $I$-ensemble, then $e_i \subseteq C_I(\cup\bar{e})$ for every $i \in I$.*
3. *If $\bar{e} \in \mathcal{F}_R{}^I$ is a perfectly-coordinated $I$-ensemble, then $\cup\bar{e} = C_I(\cup\bar{e})$.*

THEOREM B.5 ([17, 12]). *Let $R \subseteq \mathcal{R}$ and let $I \subseteq \mathbb{I}$.*

1. *For every $\psi \in \mathcal{F}_R$, the function $f_\psi^\diamondsuit : \mathcal{F}_R \to \mathcal{F}_R$ given by $x \mapsto \cap_{i \in I} \diamondsuit K_i(\psi \cap x)$ has a greatest fixed point, denoted by $C_I^\diamondsuit\psi$ — for **eventual common knowledge** of $\psi$ by $I$.*
2. *$(K_iC_I^\diamondsuit\psi)_{i \in I}$ constitutes an eventually-coordinated $I$-ensemble, for every $\psi \in \mathcal{F}_R$.*
3. *If $\bar{e} \in \mathcal{F}_R{}^I$ is an eventually-coordinated $I$-ensemble, then $e_i \subseteq K_iC_I^\diamondsuit(\cup\bar{e})$ for every $i \in I$.*
4. *If $\bar{e} \in \mathcal{F}_R{}^I$ is an eventually-coordinated $I$-ensemble, then $\cup\bar{e} \subseteq C_I^\diamondsuit(\cup\bar{e})$.*

We note that for $\varepsilon \equiv 0$, $\varepsilon$-coordination is the same as perfect coordination, and thus the following theorem also implies Theorem B.4 as a special case thereof.

THEOREM B.6 ([12]). *Let $R \subseteq \mathcal{R}$, let $I \subseteq \mathbb{I}$ and let $\varepsilon \geq 0$. For every $\psi \in \mathcal{F}_R$, denote*

$$E_I^\varepsilon(\psi) \triangleq \left\{(r,t) \in \Omega_R \;\middle|\; \begin{array}{l} \exists T \subseteq \mathbb{T} : \\ t \in T \;\&\; \sup\{T - T\} \leq \varepsilon \;\&\; \\ \forall i \in I \; \exists t' \in T : (r, t') \in K_i\psi \end{array}\right\}.$$

1. *For every $\psi \in \mathcal{F}_R$, the function $f_\psi^\varepsilon : \mathcal{F}_R \to \mathcal{F}_R$ given by $x \mapsto E_I^\varepsilon(\psi \cap x)$ has a greatest fixed point, denoted by $C_I^\varepsilon\psi$ — for **$\varepsilon$-common knowledge** of $\psi$ by $I$.*
2. *$(K_iC_I^\varepsilon\psi)_{i \in I}$ constitutes an $\varepsilon$-coordinated $I$-ensemble, for every $\psi \in \mathcal{F}_R$.*
3. *If $\bar{e} \in \mathcal{F}_R{}^I$ is an $\varepsilon$-coordinated $I$-ensemble, then $e_i \subseteq K_iC_I^\varepsilon(\cup\bar{e})$ for every $i \in I$.*
4. *If $\bar{e} \in \mathcal{F}_R{}^I$ is an $\varepsilon$-coordinated $I$-ensemble, then $\cup\bar{e} \subseteq C_I^\varepsilon(\cup\bar{e})$.*

# C. PROOFS

## C.1 Preliminaries

OBSERVATION C.1. *Let $R \subseteq \mathcal{R}$ and $i \in I$. By the positive introspection axiom, the event $K_i\psi$ is i-local for every $\psi \in \mathcal{F}_R$.*

DEFINITION C.2. *To aid the readability of the proofs below, we define $\Delta = \mathbb{Z} \cup \{\infty\}$ — the set of suprema of sets of time differences. (For every timely-coordination spec $(I, \delta)$, this is the range of $\delta$. See the definition of a timely-coordination spec in Section 4 for more details.)*[12]

OBSERVATION C.3. *By definition of $\circledcirc^{\leq \varepsilon}$,*

- $\circledcirc^{\leq \infty} = \diamondsuit$.
- $\circledcirc^{\leq 0}\psi$ means "$\psi$ has occurred, either now or in the past".
- *Additivity:* $\circledcirc^{\leq \varepsilon_1}\circledcirc^{\leq \varepsilon_2}\psi = \circledcirc^{\leq \varepsilon_1 + \varepsilon_2}\psi$ *for every* $\varepsilon_1, \varepsilon_2 \in \Delta$ *and for every* $\psi \in \mathcal{F}_R$.
- *Monotonicity:* $(\varepsilon_1 \leq \varepsilon_2 \text{ \& } \psi \subseteq \phi) \Rightarrow \circledcirc^{\leq \varepsilon_1}\psi \subseteq \circledcirc^{\leq \varepsilon_2}\phi$, *for every* $\varepsilon_1, \varepsilon_2 \in \Delta$ *and for every* $\psi, \phi \in \mathcal{F}_R$.
- $\circledcirc^{\leq \varepsilon}(\cap \Psi) \subseteq \bigcap \{\circledcirc^{\leq \varepsilon}\psi \mid \psi \in \Psi\}$, *for every* $\varepsilon \in \Delta$ *and for every set of events* $\Psi \subseteq \mathcal{F}_R$.

OBSERVATION C.4. *By definition of $\circledcirc^{\varepsilon}$, for every event $\psi \in \mathcal{F}_R$ we have:*

- $\circledcirc^{\varepsilon_1} \circledcirc^{\leq \varepsilon_2} \psi = \circledcirc^{\leq \varepsilon_1} \circledcirc^{\varepsilon_2} \psi = \circledcirc^{\leq \varepsilon_1 + \varepsilon_2}\psi$, *for every* $\varepsilon_1, \varepsilon_2 \in \Delta \setminus \{\infty\}$.
- $\circledcirc^{\varepsilon}\psi \subseteq \circledcirc^{\leq \varepsilon}\psi$, *for every $\varepsilon \in \Delta \setminus \{\infty\}$.*
- $\circledcirc^{\varepsilon}$ *commutes with intersection for every $\varepsilon \in \Delta \setminus \{\infty\}$:* $\circledcirc^{\varepsilon}(\cap \Psi) = \bigcap \{\circledcirc^{\varepsilon}\psi \mid \psi \in \Psi\}$ *for every set of events* $\Psi \subseteq \mathcal{F}_R$.

## C.2 Proofs of Propositions from Section 4

The soundness of our definition of timely common knowledge is based on the following part of Tarksi's celebrated theorem.

DEFINITION C.5 (COMPLETE LATTICE). *A lattice $L$ is called **complete** if each subset $S \subseteq L$ has both a supremum (i.e. least upper bound, denoted $\bigvee S$) and an infimum (i.e. greatest lower bound, denoted $\bigwedge S$).*

THEOREM C.6 (TARSKI [29]). *Let $L$ be a complete lattice. Every monotone function $f : L \to L$ has a greatest fixed point. Furthermore, this greatest fixed point is given by $\bigvee \{l \in L \mid l \leq f(l)\}$.*

OBSERVATION C.7. $\mathcal{F}_R^I$, *equipped with the lattice structure from Definition 4.3, constitutes a complete lattice; the supremum of every subset of $\mathcal{F}_R^I$ is given by coordinate-wise union, and its infimum — by coordinate-wise intersection.*

PROOF OF LEMMA 4.5. By monotonicity of $K_i$ for every $i \in \mathbb{I}$ and of $\circledcirc^{\leq \varepsilon}$ for every $\varepsilon \in \Delta$, we obtain that $f_\psi^\delta$ is monotone. By Observation C.7, and by Tarski's Theorem C.6, the set of fixed points of $f_\psi^\delta$ has a greatest element, which equals $\bigvee \{\bar{\xi} \in \mathcal{F}_R^I \mid \bar{\xi} \leq f_\psi^\delta(\bar{\xi})\}$. This proves both that $C_I^\delta \psi$ is well-defined (part 1 of the lemma) and the induction rule for timely common knowledge (part 2).

To prove monotonicity of $C_I^\delta$ (part 3), let $\psi, \phi \in \mathcal{F}_R$ s.t. $\psi \subseteq \phi$. Once again, by monotonicity of $K_i$ for every $i \in \mathbb{I}$, we obtain that $f_\psi^\delta(\bar{\varphi}) \leq f_\phi^\delta(\bar{\varphi})$ for every $\bar{\varphi} \in \mathcal{F}_R^I$. By substituting $\bar{\varphi} \triangleq C_I^\delta \psi$, and by definition of $C_I^\delta \psi$, we obtain $C_I^\delta \psi = f_\psi^\delta(C_I^\delta \psi) \leq f_\phi^\delta(C_I^\delta \psi)$. By directly applying the induction rule for timely common knowledge with $\bar{\xi} \triangleq C_I^\delta \psi$, we obtain that $C_I^\delta \psi \leq C_I^\delta \phi$. □

PROOF OF THEOREM 4.6. We begin the proof of part 1 by noting that for every $i \in I$, by definition $C_I^\delta \psi = f_\psi^\delta(C_I^\delta \psi)$, and therefore $(C_I^\delta \psi)_i$ is of the form $K_i(\cdots)$. Hence, by Observation C.1, $C_I^\delta \psi$ is an $I$-ensemble. Let $(i, j) \in I^2$ and $(r, t) \in (C_I^\delta \psi)_i$. By definition of $C_I^\delta$ and by the knowledge axiom,

$$(C_I^\delta \psi)_i = K_i\Big(\psi \cap \bigcap_{k \in I \setminus \{i\}} \circledcirc^{\leq \delta(i,k)}(C_I^\delta \psi)_k\Big) \subseteq$$
$$\subseteq \psi \cap \bigcap_{k \in I \setminus \{i\}} \circledcirc^{\leq \delta(i,k)}(C_I^\delta \psi)_k \subseteq \circledcirc^{\leq \delta(i,j)}(C_I^\delta \psi)_j.$$

Thus, we obtain that $(r, t) \in \circledcirc^{\leq \delta(i,j)}(C_I^\delta \psi)_j$. By definition of $\circledcirc^{\leq \delta(i,j)}$, there exists $t' \in \mathbb{T}$ such that $t' \leq t + \delta(i, j)$ and $(r, t') \in (C_I^\delta \psi)_j$, and the proof of part 1 is complete. Similarly, we have

$$(C_I^\delta \psi)_i = K_i\Big(\psi \cap \bigcap_{k \in I \setminus \{i\}} \circledcirc^{\leq \delta(i,k)}(C_I^\delta \psi)_k\Big) \subseteq$$
$$\subseteq \psi \cap \bigcap_{k \in I \setminus \{i\}} \circledcirc^{\leq \delta(i,k)}(C_I^\delta \psi)_k \subseteq \psi$$

for every $i \in I$, thus proving part 2 as well.

We move on to proving part 3. Let $\bar{e}$ be a $\delta$-coordinated $I$-ensemble s.t. $\cup \bar{e} \subseteq \psi$. First, we show that $\bar{e} \leq f_\psi^\delta(\bar{e})$. Let $i \in I$. Let $(r, t) \in e_i$ and let $j \in I \setminus \{i\}$. Since $\bar{e}$ is $\delta$-coordinated, there exists $t' \in \mathbb{T}$ s.t. $t' \leq t + \delta(i, j)$ and $(r, t') \in e_j$. By definition of $\circledcirc^{\leq \delta(i,j)}$, we therefore obtain $(r, t) \in \circledcirc^{\leq \delta(i,j)}e_j$. Thus, and since $\cup \bar{e} \subseteq \psi$, we have

$$e_i \subseteq \psi \cap \bigcap_{j \in I \setminus \{i\}} \circledcirc^{\leq \delta(i,j)}e_j.$$

By definition of an ensemble, $e_i$ is $i$-local, and thus $e_i = K_i e_i$. Hence, by monotonicity of $K_i$,

$$e_i = K_i e_i \subseteq K_i\Big(\psi \cap \bigcap_{j \in I \setminus \{i\}} \circledcirc^{\leq \delta(i,j)}e_j\Big) = \big(f_\psi^\delta(\bar{e})\big)_i.$$

By the induction rule for timely common knowledge, we thus have $\bar{e} \leq C_I^\delta \psi$, completing the proof of part 3. Part 4 follows from part 3 by setting $\psi \triangleq \cup \bar{e}$. Finally, one direction of part 5 follows from part 4 by taking the union of both sides, while the other follows by setting $\psi \triangleq \cup \bar{e}$ in part 2. □

## C.3 Proofs of Propositions from Section 5

### C.3.1 Preliminaries

In order to harness the tools of Section 4 to analyzing timely-coordinated response, we introduce some machinery relating agent responses in a protocol $P \in \mathbb{P}$ to an ensemble in the space $\Omega_{R(P,\gamma)}$ defined by the set of runs of $P$. Recall that as mentioned above, we slightly abuse notation at times when working in $\Omega_{R(P,\gamma)}$ for some protocol $P$, by writing $\phi$ to refer to $\phi \cap \Omega_{R(P,\gamma)}$.

---

[12] As noted above, we more generally define the set of time differences as $\Delta = (\mathbb{T} - \mathbb{T}) \cup \{\infty\}$. E.g. if $\mathbb{T} = \mathbb{R}_{\geq 0}$, then $\Delta = (-\infty, \infty]$.

DEFINITION C.8. *Let $\tau = (\gamma, \phi, I, \delta, \bar{\alpha})$ be a TCR and let $P \in \mathbb{P}$. We denote by $e^{P_{\bar{\alpha}}} \in \mathcal{F}_{R(P,\gamma)}{}^I$ the $I$-ensemble $e_i^{P_{\bar{\alpha}}} \triangleq \{(r,t) \in \Omega_{R(P,\gamma)} \mid i \text{ performs } \alpha_i \text{ at } (r,t) \text{ according to } P\}$, for every $i \in I$.*

OBSERVATION C.9. *Let $\tau = (\gamma, \phi, I, \delta, \bar{\alpha})$ be a TCR and let $P \in \mathbb{P}$. Since the actions of each agent $i \in I$ at each point are defined by its state at that point, it follows that $e_i^{P_{\bar{\alpha}}}$ is $i$-local, and thus $\bar{e}^{P_{\bar{\alpha}}}$ is indeed an $I$-ensemble.*

OBSERVATION C.10. *Let $\tau = (\gamma, \phi, I, \delta, \bar{\alpha})$ be a TCR. A protocol $P \in \mathbb{P}$ solves $\tau$ iff all the following hold in $\Omega_{R(P,\gamma)}$:*

- *$e_i^{P_{\bar{\alpha}}}$ occurs at most once during each run $r \in R(P,\gamma)$, for every agent $i \in I$.*
- *$\bar{e}^{P_{\bar{\alpha}}}$ is $\delta$-coordinated.*
- *$\cup \bar{e}^{P_{\bar{\alpha}}} \subseteq \odot^{\leq 0}\phi$. (I.e. $\phi$ must occur before or when any response does.)*
- *$\phi \subseteq \diamondsuit e_i^{P_{\bar{\alpha}}}$, for every $i \in I$. (I.e. all responses must occur at some point along any $\phi$-triggered run.)*

OBSERVATION C.11. *Let $\tau = (\gamma, \phi, I, \delta, \bar{\alpha})$ be a TCR. A protocol $P \in \mathbb{P}$ is a time-optimal solution to $\tau$ iff it both solves $\tau$ and for every protocol $P'$ solving $\tau$ that is run-equivalent to $P$, we have $e_i^{P'_{\bar{\alpha}}} \subseteq \odot^{\leq 0} e_i^{P_{\bar{\alpha}}}$ in $\Omega_{R(P_{-\bar{\alpha}},\gamma)}$ for every $i \in I$.*

### C.3.2 Proofs

PROOF OF COROLLARY 5.2. We must show that under the conditions of the corollary, $\bar{e}^{P_{\bar{\alpha}}} \leq C_I^\delta(\odot^{\leq 0}\phi)$ holds in $\Omega_{R(P,\gamma)}{}^I$. Since $P$ solves $\tau$, by Observation C.10 we have both that $\bar{e}^{P_{\bar{\alpha}}}$ is $\delta$-coordinated and that $\cup \bar{e}^{P_{\bar{\alpha}}} \subseteq \odot^{\leq 0}\phi$. Thus, by Theorem 4.6(3), we obtain $\bar{e}^{P_{\bar{\alpha}}} \leq C_I^\delta(\odot^{\leq 0}\phi)$, as required. □

The following somewhat technically-phrased lemma lies at the heart of Corollaries C.13 and 5.3, whose proofs follow below.

LEMMA C.12. *Let $\tau = (\gamma, \phi, I, \delta, \bar{\alpha})$ be a TCR and let $P_{-\bar{\alpha}}$ be a non-response component of a protocol such that $\phi \subseteq \diamondsuit \bigl(C_I^\delta(\odot^{\leq 0}\phi)\bigr)_i$ holds in $\Omega_{R(P_{-\bar{\alpha}},\gamma)}$ for some $i \in I$. The protocol $P = (P_{-\bar{\alpha}}, P_{\bar{\alpha}})$ s.t. in $P_{\bar{\alpha}}$ each $i \in I$ responds at the first instant at which $\bigl(C_I^\delta(\odot^{\leq 0}\phi)\bigr)_i$ holds (in $\Omega_{R(P_{-\bar{\alpha}},\gamma)}$), is a time-optimal solution for $\tau$.*

PROOF. We note that by Theorem 4.6(1), $\bigl(C_I^\delta(\odot^{\leq 0}\phi)\bigr)_j$ is $j$-local for every $j \in I$, and thus $P_{\bar{\alpha}}$ is well-defined.[13] We now show that $P$ solves $\tau$ by showing that it satisfies all four conditions of Observation C.10.

---
[13] In some runs of certain contexts under a continuous-time model, the set of times at which $\bigl(C_I^\delta(\odot^{\leq 0}\phi)\bigr)_j$ holds does not attain its infimum value, and thus "$\bigl(C_I^\delta(\odot^{\leq 0}\phi)\bigr)_j$ holds for the first time" is not necessarily a $j$-local event. To accommodate such cases, we may adapt the response component $P_{\bar{\alpha}}$ s.t. each $j \in I$ responds exactly 1 time unit after the infimum of times at which $\bigl(C_I^\delta(\odot^{\leq 0}\phi)\bigr)_j$ holds. (It is straightforward to show that this is indeed a $j$-local event). The proof is easily adaptable to both show that this definition yields a solution for $\tau$ and to prove that in such pathological cases, no time-optimal solution for $\tau$ exists.

By definition of $P_{\bar{\alpha}}$, for each $j \in I$ the event $e_j^{P_{\bar{\alpha}}}$ occurs at most once during each $r \in R(P,\gamma)$. Let $(j,k) \in I^2$ and let $(r,t) \in e_j^{P_{\bar{\alpha}}}$. By definition of $P_{\bar{\alpha}}$, we have that $(r,t) \in \bigl(C_I^\delta(\odot^{\leq 0}\phi)\bigr)_j$. By Theorem 4.6(1), $C_I^\delta(\odot^{\leq 0}\phi)$ is a $\delta$-coordinated ensemble, and thus there exists $t' \leq t + \delta(j,k)$ s.t. $(r,t') \in \bigl(C_I^\delta(\odot^{\leq 0}\phi)\bigr)_k$. By definition of $P_{\bar{\alpha}}$, there exists $t'' \leq t'$ s.t. $(r,t'') \in e_k^{P_{\bar{\alpha}}}$. As $t'' \leq t' \leq t + \delta(j,k)$, we obtain that $\bar{e}^{P_{\bar{\alpha}}}$ is $\delta$-coordinated.

Let $j \in I$. By Observation C.3 (monotonicity), we conclude that $e_i^{P_{\bar{\alpha}}} \subseteq \odot^{\leq \delta(i,j)} e_j^{P_{\bar{\alpha}}} \subseteq \diamondsuit e_j^{P_{\bar{\alpha}}}$. By definition of $P_{\bar{\alpha}}$, we have $\bigl(C_I^\delta(\odot^{\leq 0}\phi)\bigr)_i \subseteq \odot^{\leq 0} e_i^{P_{\bar{\alpha}}}$. By both of these, by the conditions of the lemma, and once again by Observation C.3 (monotonicity), we obtain $\phi \subseteq \diamondsuit \bigl(C_I^\delta(\odot^{\leq 0}\phi)\bigr)_i \subseteq \diamondsuit \odot^{\leq 0} e_i^{P_{\bar{\alpha}}} \subseteq \diamondsuit \odot^{\leq 0} e_j^{P_{\bar{\alpha}}} = \diamondsuit e_j^{P_{\bar{\alpha}}}$. Finally, by definition of $P_{\bar{\alpha}}$ and by Theorem 4.6(2), we have $\cup \bar{e}^{P_{\bar{\alpha}}} \subseteq \cup C_I^\delta(\odot^{\leq 0}\phi) \subseteq \odot^{\leq 0}\phi$, thus completing the proof of $P$ solving $\tau$.

We move on to show that $P$ constitutes a time-optimal solution to $\tau$. Let $P' = (P_{-\bar{\alpha}}, P'_{\bar{\alpha}})$ be a protocol solving $\tau$ that is run-equivalent to $P$. Let $j \in I$. By Corollary 5.2, we have $e_j^{P'_{\bar{\alpha}}} \subseteq \bigl(C_I^\delta(\odot^{\leq 0}\phi)\bigr)_j$. By definition of $P_{\bar{\alpha}}$, we have $\bigl(C_I^\delta(\odot^{\leq 0}\phi)\bigr)_j \subseteq \odot^{\leq 0} e_j^{P_{\bar{\alpha}}}$. We combine these to obtain $e_j^{P'_{\bar{\alpha}}} \subseteq \odot^{\leq 0} e_j^{P_{\bar{\alpha}}}$, and thus, by Observation C.11, the proof is complete. □

COROLLARY C.13. *Let $\tau = (\gamma, \phi, I, \delta, \bar{\alpha})$ be a TCR and let $P \in \mathbb{P}$. The following are equivalent:*

1. *$P$ is run-equivalent to a protocol that solves $\tau$.*
2. *$\phi \subseteq \diamondsuit \bigl(C_I^\delta(\odot^{\leq 0}\phi)\bigr)_i$ in $\Omega_{R(P,\gamma)}$, for every $i \in I$.*
3. *$\phi \subseteq \diamondsuit \bigl(C_I^\delta(\odot^{\leq 0}\phi)\bigr)_i$ in $\Omega_{R(P,\gamma)}$, for some $i \in I$.*

PROOF.
$1 \Rightarrow 2$: Let $i \in I$. Let $P'$ be a protocol solving $\tau$ that is run-equivalent to $P$. Recall that $\Omega_{R(P',\gamma)} \simeq \Omega_{R(P,\gamma)}$. By Observation C.10, we have $\phi \subseteq \diamondsuit e_i^{P'_{\bar{\alpha}}}$. By Corollary 5.2, we have $e_i^{P'_{\bar{\alpha}}} \subseteq \bigl(C_I^\delta(\odot^{\leq 0}\phi)\bigr)_i$. We combine these two with Observation C.3 (monotonicity) to obtain $\phi \subseteq \diamondsuit \bigl(C_I^\delta(\odot^{\leq 0}\phi)\bigr)_i$.

$2 \Rightarrow 3$: Immediate.

$3 \Rightarrow 1$: Follows immediately from Lemma C.12, since $\Omega_{R(P,\gamma)} \simeq \Omega_{R(P_{-\bar{\alpha}},\gamma)}$. □

PROOF OF COROLLARY 5.3. By Corollary C.13($1 \Rightarrow 2$), we have $\phi \subseteq \diamondsuit \bigl(C_I^\delta(\odot^{\leq 0}\phi)\bigr)_i$ holding in $\Omega_{R(P,\gamma)} \simeq \Omega_{R(P_{-\bar{\alpha}},\gamma)}$. By Lemma C.12, the proof is complete. □

## C.4 From Fixed-Point Definition to Nested-Knowledge Definition

### C.4.1 Definitions and Propositions

In order to precisely phrase our nested-knowledge characterisation of timely common knowledge, we first introduce an additional definition.[14]

---
[14] As our notation $\mathcal{P}(G_\delta)$ may suggest, this is in fact the set of paths in a directed graph $G_\delta$ having $I$ as vertices and with edges wherever $\delta < \infty$. For an in-depth graph-theoretic study of $G_\delta$ and of its elaborate relation to tuples of $\delta$-coordinated timestamps, we refer the reader to [15] or to [16, Chapter 5]. For a study of the connection between the graph-theoretic properties of $G_\delta$ and the required delivery guarantees required to solve a TCR, we refer the reader to [16, Chapter 9].

DEFINITION C.14. *Let $I$ be a set and let $\delta : I^{\bar{2}} \to \Delta$. We define*

$$\mathcal{P}(G_\delta) \triangleq \{(i_1,\ldots,i_n) \in I^{\bar{*}} \mid \forall m \in [n-1] : \delta(i_m,i_{m+1}) < \infty\}.$$

EXAMPLE C.15. *By the above definition, if $I = \{i,j\}$, then every element of $\mathcal{P}(G_\delta)$ is either $\underbrace{(i,j,i,j,i,j,\ldots)}_{n}$ or $\underbrace{(j,i,j,i,j,i,\ldots)}_{n}$, for some $n \in \mathbb{N}$. (If $|I| > 2$, then $\mathcal{P}(G_\delta)$ is much richer.)*

Second, we present a variation of a definition from [12, Chapter 4], which we utilize in this section.

DEFINITION C.16 (PERFECT RECALL).
*A system $R \subseteq \mathcal{R}$ is said to exhibit **perfect recall** if for every $r \in R$, for every $i \in \mathbb{I}$ and for every $t \in \mathbb{T}$, the state of $i$ at $t$ in $r$ uniquely determines the set $\{r_i(t') \mid t' \in \mathbb{T} \setminus [t,\infty)\}$ of states of $i$ in $r$ prior to $t$.*

OBSERVATION C.17. *If $P_\gamma^{\text{fip}}$ is a full-information protocol in a context $\gamma$, then $R(P_\gamma^{\text{fip}},\gamma)$ exhibits perfect recall.*

Third, we present a definition based upon [12, Chapter 4] and some basic properties thereof.

DEFINITION C.18 (STABILITY). *Let $R \subseteq \mathcal{R}$. An event $\psi \in \mathcal{F}_R$ is said to be **stable** if once $\psi$ holds at some time during a run $r \in \mathcal{R}$, it continues to hold for the duration of $r$. Formally, using our notation, $\psi$ is stable iff $\psi = \circledcirc^{\leq 0}\psi$.*

OBSERVATION C.19. *By Definition C.18,*
- *By Observation C.3 (additivity), $\circledcirc^{\leq 0}$ is idempotent. Thus, $\circledcirc^{\leq 0}\phi$ is a stable event for every $\phi \in \mathcal{F}_R$.*
- *$\psi \cap \phi$ is a stable event for any two stable events $\psi, \phi \in \mathcal{F}_R$.*

Indeed, since $\circledcirc^{\leq 0}\phi$ is stable for every $\phi$, we do not lose much in the perspective of Section 5 if we restrict our study to timely common knowledge of stable events. We can now precisely phrase our constructive characterisation of timely common knowledge. See the following sections for a proof and a discussion of the various requirements of the following theorem.

THEOREM C.20. *Let $(I,\delta)$ be a timely-coordination spec, let $R \subseteq \mathcal{R}$ be a system exhibiting perfect recall and let $\psi \in \mathcal{F}_R$ be a stable event. Assume, furthermore, that either of the following holds:*
1. *$\delta < \infty$.*
2. *$R = R(P,\gamma)$, for some protocol $P$ and context $\gamma$ s.t. $P$ either solves $(\gamma,\psi,I,\delta,\bar{\alpha})$ for some $\bar{\alpha}$, or is run-equivalent to a protocol that does.*

*For every $i \in I$,*

$$(C_I^\delta \psi)_i = \bigcap_{(i,i_2,\ldots,i_n) \in \mathcal{P}(G_\delta)} K_i \circledcirc^{\delta(i,i_2)} K_{i_2} \circledcirc^{\delta(i_2,i_3)} K_{i_3} \cdots \circledcirc^{\delta(i_{n-1},i_n)} K_{i_n} \psi \quad (2)$$

*holds in $\Omega_R$.*

OBSERVATION C.21. *By Observation C.17, and since it is straitforward to show that a TCR is solvable iff it is solvable by a full-information protocol, condition 2 of Theorem C.20 is met if $R = R(P_\gamma^{\text{fip}},\gamma)$, for a context $\gamma$ admitting a full-information protocol $P_\gamma^{\text{fip}}$ s.t. $(\gamma,\psi,I,\delta,\bar{\alpha})$ is solvable (by some protocol) for some $\bar{\alpha}$.*

COROLLARY C.22. *The time-optimal solution from Corollary 5.3, under (any of) the conditions of Theorem C.20 (with regard to $R \triangleq R(P_{-\bar{\alpha}},\gamma)$ and $\psi \triangleq \circledcirc^{\leq 0}\phi$), is for each agent $i \in I$ to respond at the first instant at which all nested-knowledge formulae of the form*

$$K_i \circledcirc^{\delta(i,i_2)} K_{i_2} \circledcirc^{\delta(i_2,i_3)} \cdots K_{i_{n-1}} \circledcirc^{\delta(i_{n-1},i_n)} K_{i_n} \circledcirc^{\leq 0} \phi$$

*hold (in $\Omega_{R(P_{-\bar{\alpha}},\gamma)}$) for all $(i,i_2,\ldots,i_n) \in \mathcal{P}(G_\delta)$.*

### C.4.2 Background

In order to prove Theorem C.20, we perform an analysis of timely common knowledge of stable events. For reasons that will soon be apparent, we conduct this analysis under the assumption of perfect recall. To make our analysis somewhat cleaner and more generic, we first aim to distill the property of sets of runs exhibiting perfect recall that is of interest to us, namely that in such sets of runs, knowledge of a stable event is itself stable. The following is given in [12, Exercise 4.18(b)], and its proof follows directly from the definitions of stability and of knowledge.

CLAIM C.23. *Let $R \subseteq \mathcal{R}$ be a system exhibiting perfect recall and let $\psi \in \mathcal{F}_R$. If $\psi$ is stable, then $K_i\psi$ is stable as well, for every $i \in \mathbb{I}$.*

### C.4.3 Proof

Returning to our results and working toward proving Theorem C.20, we first derive a stability property for timely common knowledge (given in Claim C.25.)

CLAIM C.24. *Let $R \subseteq \mathcal{R}$ be a system exhibiting perfect recall. For every event $\psi \in \mathcal{F}_R$ and for every agent $i \in I$, it holds that $\circledcirc^{\leq 0} K_i \psi \subseteq K_i \circledcirc^{\leq 0} \psi$.*

PROOF. By Observation C.3, we have $\psi \subseteq \circledcirc^{\leq 0}\psi$. Thus, by monotonicity of $\circledcirc^{\leq 0}$ and of $K_i$, we have $\circledcirc^{\leq 0} K_i \psi \subseteq \circledcirc^{\leq 0} K_i \circledcirc^{\leq 0} \psi$. By Observation C.19, $\circledcirc^{\leq 0} \psi$ is stable, and therefore, by Claim C.23, $K_i \circledcirc^{\leq 0} \psi$ is stable as well, and thus equals $\circledcirc^{\leq 0} K_i \circledcirc^{\leq 0} \psi$, by applying Observation C.19 once more. We combine all these to obtain $\circledcirc^{\leq 0} K_i \psi \subseteq \circledcirc^{\leq 0} K_i \circledcirc^{\leq 0} \psi = K_i \circledcirc^{\leq 0} \psi$, as required. □

CLAIM C.25. *Let $(I,\delta)$ be a timely-coordination spec and let $R \subseteq \mathcal{R}$ be a set of runs exhibiting perfect recall. For every stable $\psi \in \mathcal{F}_R$, all coordinates of $C_I^\delta \psi$ are stable.*

PROOF. Let $i \in I$. By Definition C.18 and by Observation C.3, it is enough to show that $\circledcirc^{\leq 0}(C_I^\delta \psi)_i \subseteq (C_I^\delta \psi)_i$. Indeed, we have

$$\circledcirc^{\leq 0}(C_I^\delta \psi)_i = \qquad \text{by definition of } C_I^\delta$$

$$= \circledcirc^{\leq 0} K_i \Big(\psi \cap \bigcap_{j \in I\setminus\{i\}} \circledcirc^{\leq \delta(i,j)}(C_I^\delta \psi)_j\Big) \subseteq$$

$$\text{by Claim C.24}$$

$$\subseteq K_i \circledcirc^{\leq 0} \Big(\psi \cap \bigcap_{j \in I\setminus\{i\}} \circledcirc^{\leq \delta(i,j)}(C_I^\delta \psi)_j\Big) \subseteq$$

$$\text{by Observation C.3}$$

$$\subseteq K_i \Big(\circledcirc^{\leq 0}\psi \cap \bigcap_{j \in I\setminus\{i\}} \circledcirc^{\leq 0} \circledcirc^{\leq \delta(i,j)}(C_I^\delta \psi)_j\Big) \subseteq$$

$$\text{by Observation C.3 (additivity)}$$

$$\subseteq K_i\Big(\circledcirc^{\leq 0}\psi \cap \bigcap_{j\in I\setminus\{i\}} \circledcirc^{\leq \delta(i,j)}(C_I^\delta\psi)_j\Big) \subseteq$$

by stability of $\psi$

$$\subseteq K_i\Big(\psi \cap \bigcap_{j\in I\setminus\{i\}} \circledcirc^{\leq \delta(i,j)}(C_I^\delta\psi)_j\Big) =$$

by definition of $C_I^\delta$

$$= (C_I^\delta\psi)_i.$$

□

Claims C.23 and C.25 lead us to consider, for stable $\psi$ and given perfect recall, a slightly different definition for $f_\psi^\delta$ than the one given in Definition 4.4. This modified version of $f_\psi^\delta$, which we denote by $g_\psi^\delta$, differs by the use of $\circledcirc^{\delta(i,j)}$ in lieu of $\circledcirc^{\leq\delta(i,j)}$, and by not intersecting over eventual knowledge requirements.

DEFINITION C.26. *Let $(I,\delta)$ be a timely-coordination spec and let $R \subseteq \mathcal{R}$. For each $\psi \in \mathcal{F}_R$, we define a function $g_\psi^\delta : \mathcal{F}_R^I \to \mathcal{F}_R^I$ by*

$$g_\psi^\delta: \quad (x_i)_{i\in I} \;\mapsto\; \left(K_i\Big(\psi \cap \bigcap_{\substack{j\in I\setminus\{i\}\\ \delta(i,j)<\infty}} \circledcirc^{\delta(i,j)} x_j\Big)\right)_{i\in I},$$

*and denote its greatest fixed point by $\mathcal{C}_I^\delta\psi$.*

Using an argument completely analogous to the proof of Lemma 4.5, it may be shown that $\mathcal{C}_I^\delta\psi$ is well-defined. Furthermore, the same argument shows that $\mathcal{C}_I^\delta$ also satisfies the obvious analogues of the induction rule (with regard to $g_\psi^\delta$) and of the monotonicity property from Lemma 4.5.

We now present a key observation, which stands at the heart of our proof of Theorem C.20. While, even in the presence of perfect recall and when $\psi$ is stable, $g_\psi^\delta \ne f_\psi^\delta$ (e.g. when applied to certain unstable events), it so happens that under certain conditions, the greatest fixed points of both of these functions coincide.

LEMMA C.27. *Let $(I,\delta)$ be a timely-coordination spec, let $R \subseteq \mathcal{R}$ be a set of runs exhibiting perfect recall and let $\psi \in \mathcal{F}_R$. Furthermore, assume that either $\psi \subseteq \Diamond(C_I^\delta\psi)_i$ for every $i \in I$, or $\delta < \infty$. If $\psi$ is stable, then $\mathcal{C}_I^\delta\psi = C_I^\delta\psi$.*

PROOF.

$\geq$: For every $i \in I$, we have

$$(C_I^\delta\psi)_i = \qquad \text{by definition of } C_I^\delta$$

$$= K_i\Big(\psi \cap \bigcap_{j\in I\setminus\{i\}} \circledcirc^{\leq \delta(i,j)}(C_I^\delta\psi)_j\Big) \subseteq$$

intersecting over fewer events

$$\subseteq K_i\Big(\psi \cap \bigcap_{\substack{j\in I\setminus\{i\}\\ \delta(i,j)<\infty}} \circledcirc^{\leq \delta(i,j)}(C_I^\delta\psi)_j\Big) =$$

by Observation C.4

$$= K_i\Big(\psi \cap \bigcap_{\substack{j\in I\setminus\{i\}\\ \delta(i,j)<\infty}} \circledcirc^{\delta(i,j)} \circledcirc^{\leq 0}(C_I^\delta\psi)_j\Big) =$$

by Claim C.25

$$= K_i\Big(\psi \cap \bigcap_{\substack{j\in I\setminus\{i\}\\ \delta(i,j)<\infty}} \circledcirc^{\delta(i,j)}(C_I^\delta\psi)_j\Big) =$$

by definition of $g_\psi^\delta$

$$= \big(g_\psi^\delta(C_I^\delta\psi)\big)_i.$$

Thus, by the induction rule for $\mathcal{C}_I^\delta$ and for $g_\psi^\delta$, we obtain $C_I^\delta\psi \leq \mathcal{C}_I^\delta\psi$, as required.

$\leq$: For every $i \in I$, by monotonicity of $K_i$ we have

$$(\mathcal{C}_I^\delta\psi)_i = \qquad \text{by definition of } \mathcal{C}_I^\delta$$

$$= K_i\Big(\psi \cap \bigcap_{\substack{j\in I\setminus\{i\}\\ \delta(i,j)<\infty}} \circledcirc^{\delta(i,j)}(\mathcal{C}_I^\delta\psi)_j\Big) \subseteq$$

by Observation C.4

$$\subseteq K_i\Big(\psi \cap \bigcap_{\substack{j\in I\setminus\{i\}\\ \delta(i,j)<\infty}} \circledcirc^{\leq\delta(i,j)}(\mathcal{C}_I^\delta\psi)_j\Big) \subseteq$$

as $\psi \subseteq \Diamond(\mathcal{C}_I^\delta\psi)_j$ for every $j \in I$
(expression unchanged if $\delta < \infty$)

$$\subseteq K_i\Big(\psi \cap \bigcap_{\substack{j\in I\setminus\{i\}\\ \delta(i,j)=\infty}} \Diamond(\mathcal{C}_I^\delta\psi)_j \cap \bigcap_{\substack{j\in I\setminus\{i\}\\ \delta(i,j)<\infty}} \circledcirc^{\leq\delta(i,j)}(\mathcal{C}_I^\delta\psi)_j\Big) \subseteq$$

by the other direction ($\geq$) of this proof, and by monotonicity of $\Diamond$

$$\subseteq K_i\Big(\psi \cap \bigcap_{\substack{j\in I\setminus\{i\}\\ \delta(i,j)=\infty}} \Diamond(C_I^\delta\psi)_j \cap \bigcap_{\substack{j\in I\setminus\{i\}\\ \delta(i,j)<\infty}} \circledcirc^{\leq\delta(i,j)}(C_I^\delta\psi)_j\Big) =$$

as $\Diamond = \circledcirc^{\leq\infty}$

$$= K_i\Big(\psi \cap \bigcap_{j\in I\setminus\{i\}} \circledcirc^{\leq\delta(i,j)}(C_I^\delta\psi)_j\Big) =$$

by definition of $f_\psi^\delta$

$$= \big(f_\psi^\delta(C_I^\delta\psi)\big)_i.$$

Thus, by the induction rule for timely common knowledge, we have $\mathcal{C}_I^\delta\psi \leq C_I^\delta\psi$.[15] □

One may wonder why we have worked so hard, and added the additional assumption of perfect recall (among others), to obtain $C_I^\delta\psi$, under the above assumptions, as a fixed point of $g_\psi^\delta$ rather than of $f_\psi^\delta$. The answer is simple: $g_\psi^\delta$ commutes with the meet operation, while $f_\psi^\delta$ does not. (Moreover, as a result, $g_\psi^\delta$ is downward-continuous while $f_\psi^\delta$, even in a discrete-time model, is generally not.) This fact paves our way toward proving Theorem C.20.

---

[15] It should be noted that we could have saved ourselves some hardship in this direction of the proof, by not intersecting over eventual knowledge requirements when defining $f_\psi^\delta$. While this would still have allowed us to obtain some of our main results, such as Corollary 5.3, in this case many of our other results regarding timely common knowledge would have required the additional assumption that $\psi \subseteq \Diamond(C_I^\delta\psi)_i$, reducing from their generality and usefulness. The added strength of the approach we have chosen presents itself not only while discussing eventual common knowledge in Appendix D.2, but in other settings [16, Section 9.3] as well.

PROOF OF THEOREM C.20. The following proof applies to prove both parts of the theorem. As $\psi$ is stable and $R$ exhibits perfect recall, by Lemma C.27,[16] we obtain $C_I^\delta \psi = \mathcal{C}_I^\delta \psi$.

It is easy to verify that $g_\psi^\delta$ commutes with both finite and infinite meet. Thus, it is downward-continuous and by a well-known theorem popularly referred to as Kleene's fixed-point theorem[17], we obtain

$$\mathcal{C}_I^\delta \psi \;=\; \bigwedge_{n \in \mathbb{N}} (g_\psi^\delta)^n \, (\Omega_R{}^I). \qquad (3)$$

Since $\circledcirc^\varepsilon$ commutes with intersection for every $\varepsilon \in \Delta$, and $K_i$ commutes with intersection for every $i \in I$, we thus obtain, for every $i \in I$, that

$(C_I^\delta \psi)_i =$

$= \bigcap_{n \in \mathbb{N}} \left( (g_\psi^\delta)^n \, (\Omega_R{}^I) \right)_i =$

$= K_i \psi \,\cap\, K_i \Big( \psi \cap \bigcap_{\substack{i_2 \in I \setminus \{i\} \\ \delta(i, i_2) < \infty}} \circledcirc^{\delta(i, i_2)} K_{i_2} \psi \Big) \cap$

$\cap\, K_i \bigg( \psi \cap \bigcap_{\substack{i_2 \in I \setminus \{i\} \\ \delta(i, i_2) < \infty}} \circledcirc^{\delta(i, i_2)} K_{i_2} \Big( \psi \cap \bigcap_{\substack{i_3 \in I \setminus \{i_2\} \\ \delta(i_2, i_3) < \infty}} \circledcirc^{\delta(i_2, i_3)} K_{i_3} \psi \Big) \bigg) \cap$

$\cap\, \cdots \;=$

$= K_i \psi \,\cap\, K_i \Big( \bigcap_{\substack{i_2 \in I \setminus \{i\} \\ \delta(i, i_2) < \infty}} \circledcirc^{\delta(i, i_2)} K_{i_2} \psi \Big) \cap$

$\cap\, K_i \bigg( \bigcap_{\substack{i_2 \in I \setminus \{i\} \\ \delta(i, i_2) < \infty}} \circledcirc^{\delta(i, i_2)} K_{i_2} \Big( \bigcap_{\substack{i_3 \in I \setminus \{i_2\} \\ \delta(i_2, i_3) < \infty}} \circledcirc^{\delta(i_2, i_3)} K_{i_3} \psi \Big) \bigg) \cap$

$\cap\, \cdots \;=$

$= \bigcap_{(i, i_2, \ldots, i_n) \in \mathcal{P}(G_\delta)} K_i \circledcirc^{\delta(i, i_2)} K_{i_2} \circledcirc^{\delta(i_2, i_3)} K_{i_3} \cdots \circledcirc^{\delta(i_{n-1}, i_n)} K_{i_n} \psi.$

□

### C.4.4 Discussion

Theorem C.20, which we have just proved, hinges on quite a few conditions, especially when $\delta \not< \infty$. The two conditions that are required even when $\delta < \infty$, namely perfect recall and stability of $\psi$, allow us to define $g_\psi^\delta$ using $\circledcirc^{\delta(i,j)}$ instead of $\circledcirc^{\leq \delta(i,j)}$. Without this modification, $g_\psi^\delta$ would not commute with intersection, resulting, instead of (2), in

$$\bigcap_{(i, i_2, \ldots, i_n) \in \mathcal{P}(G_\delta)} K_i \Big( \psi \cap \circledcirc^{\delta(i, i_2)} K_{i_2} \big( \cdots \psi \cap \circledcirc^{\delta(i_{n-1}, i_n)} K_{i_n}(\psi) \cdots \big) \Big).$$

---

[16] For the proof given condition 2, at this point we also use the fact that by Corollary C.13, we have that $\psi \subseteq \diamondsuit \big( C_I^\delta (\circledcirc^{\leq 0} \psi) \big)_i = \diamondsuit (C_I^\delta \psi)_i$ for every $i \in I$.

[17] See [23] for an investigation of the origins of this theorem, see [20, p. 348] for Kleene's first recursion theorem and for its proof that implies this theorem, and see [21] or [1, Theorem 1.2.14] for a statement of this theorem in terms of lattices, continuity and greatest fixed points.

When $\delta \not< \infty$, it is condition 2 of Theorem C.20 that allows us to define $g_\psi^\delta$ without intersecting over eventual knowledge requirements. Without this modification, $g_\psi^\delta$ would not be downward-continuous. (This would also have been the case, had $g_\psi^\delta$ been defined using $\circledcirc^{\leq \delta(i,j)}$ instead of $\circledcirc^{\delta(i,j)}$ when under a continuous-time model.) Without downward continuity of $g_\psi^\delta$, Kleene's fixed-point theorem could not have been utilized, forcing us to go beyond the "$\omega$'th power" of $g_\psi^\delta$ in the r.h.s. of (3), to a greater ordinal power thereof [1, Theorem 1.2.11]. Incidentally, this may be viewed as a concrete example, of sorts, of Barwise's statement in [3] regarding various definitions of common knowledge, according to which in some models, taking only the intersection of finite approximations (i.e. only the results of finitely-many iterations of the relevant function $f$, starting from the top of the lattice) yields a weaker state of knowledge than the fixed-point of $f$, which is equivalent to taking the intersection of all (i.e. including transfinite) approximations.

We conclude this section with an observation. For certain $\delta$ functions, $\mathcal{P}(G_\delta)$ is finite,[18] and thus the intersection in (2) is finite. (See the discussion of ordered response in Appendix D.1 for an example.) This observation may seem, at first glance, to clash with the infinitary nature of fixed points in general, and of greatest fixed points in particular. It is worthwhile to note that what reconciles these is that in this case, $(g_\psi^\delta)^{|I|}$ is constant and therefore its value, which is a finite intersection of nested-knowledge events, is its only fixed point, and thus its greatest fixed point, and hence the greatest fixed point of $g_\psi^\delta$ as well. Furthermore, by Corollary C.13, solvability of $\tau$ implies that $\psi \subseteq \diamondsuit (C_I^\delta \psi)_i$ for every $i \in I$ and thus, as noted above, Corollaries 5.2 and 5.3 would have still held had we defined $f_\psi^\delta$ without intersecting over eventual-knowledge requirements (i.e. similarly to $g_\psi^\delta$, but using $\circledcirc^{\leq \delta(i,j)}$ in lieu of $\circledcirc^{\delta(i,j)}$). In this case, the function $(f_\psi^\delta)^{|I|}$ would have also been constant, and thus, similarly, its value would have been its greatest fixed point, and thus the greatest fixed point of $f_\psi^\delta$ as well.

## D. COMPARISON TO, AND DERIVATION OF PREVIOUS RESULTS

In this section, we show how some previously-known results may be derived from the novel results we have introduced in this paper.

### D.1 Response Problems

In this section, we survey the response problems defined and studied by Ben-Zvi and Moses [5, 4, 6, 7], and their knowledge-theoretic results for these problems. We reformulate these problems, their results, and the associated definitions to match our notation, and show how our definitions and results from Section 5 extend each one of these, even though the tools used to derive our results are vastly different than their tools. This provides us with a "sanity check" of sorts, verifying that we have not committed the sin of generalizing our tools to the extent of weakening the results they yield for simple cases.

The first, most-basic response problem defined in [5] is that of *ordered response*. In this problem, finitely many agents $I = \{i_m\}_{m=1}^n$ must respond to an event in a pre-defined order: $i_{m+1}$ may not respond before $i_m$ does. Using our notation, this is a special case of timely-coordinated response, for

$$\delta(i_k, i_l) \triangleq \begin{cases} 0 & k = l+1 \\ \infty & \text{otherwise.} \end{cases} \quad (4)$$

---

[18] This happens iff both $|I| < \infty$ and $G_\delta$ has only trivial (i.e. singleton) strongly connected components.

For this problem, they have shown that whenever an agent $i_m$ responds in a solving protocol exhibiting perfect recall, it holds that

$$K_m K_{m-1} \cdots K_1 \circledcirc^{\leq 0} \phi, \quad (5)$$

and that in a full-information protocol, for each $i_m$ to respond as soon as (5) holds constitutes what we have defined as a time-optimal solution.[19]

By Theorem C.20, we have, for $\delta$ as defined in (4), that when $\tau$ is solvable and in the presence of perfect recall (e.g. in a full-information protocol),

$$\begin{aligned} \left(C_I^\delta(\circledcirc^{\leq 0}\phi)\right)_m &= \bigcap_{\substack{k \in \mathbb{N} \\ 1 \leq i_1 < \cdots < i_k \leq m}} K_{i_k} \cdots K_{i_1} \circledcirc^{\leq 0} \phi = \\ &= K_m K_{m-1} \cdots K_1 \circledcirc^{\leq 0} \phi. \end{aligned}$$

Thus, for ordered response, Corollaries 5.2 and 5.3 reduce to the above results.

The second problem presented in [5, 4] is a variant of the firing squad problem [27, 10] called *simultaneous response*. In this problem, all agents $I$ must respond to an event simultaneously. Using our notation, this is a special case of timely-coordinated response, for $\delta \equiv 0$. For this problem, they have shown that $C_I \circledcirc^{\leq 0} \phi$ is the associated state of knowledge, in the same sense as above, i.e. when the agents respond (in a solving protocol exhibiting perfect recall), they share common knowledge of the fact that $\phi$ has occurred, and for each agent to respond as soon as she knows that common knowledge of $\circledcirc^{\leq 0}\phi$ has been attained constitutes a time-optimal solution (for a full-information protocol, when the problem is solvable). Once again, Corollaries 5.2 and 5.3 reduce to the above results under the above assumptions, since by Theorem C.20 and by Observation A.4, we have $\left(C_I^\delta(\circledcirc^{\leq 0}\phi)\right)_i = K_i C_I \circledcirc^{\leq 0} \phi = C_I \circledcirc^{\leq 0} \phi$, for $\delta \equiv 0$.

The third and last problem presented in [4] is a generalization of both ordered response and simultaneous response, called *ordered joint response*. In this problem, the agents are partitioned into pairwise-disjoint sets $I = \bigcup_{m=1}^n I_m$, and the agents in each such set must respond simultaneously, s.t. the agents in a set $I_{m+1}$ may not respond before the agents in $I_m$ do. Under our notation, this is a special case of timely-coordinated response, for

$$\delta(i,j) \triangleq \begin{cases} 0 & \begin{aligned} &\exists k \in [n]: \{i,j\} \subseteq I_k \quad \text{or} \\ &\exists k \in [n-1]: i \in I_{k+1} \ \& \ j \in I_k \end{aligned} \\ \infty & \text{otherwise.} \end{cases} \quad (6)$$

For this problem, they have shown that the associated state of knowledge, in the above sense, for an agent $i \in I_m$, is $C_{I_m} C_{I_{m-1}} \cdots C_{I_1} \circledcirc^{\leq 0} \phi$.

By Theorem C.20, by Observation A.4 and by $K_j$ commuting with intersection for every $j \in I$, we have, in this case, for $\delta$ as defined in (6) and for $i \in I_m$,

$$\left(C_I^\delta(\circledcirc^{\leq 0}\phi)\right)_i =$$

---
[19] Throughout their analysis, Ben-Zvi and Moses implicitly assume that the problems they consider are solvable.

$$\begin{aligned} &= \bigcap_{(i,i_2,\ldots,i_n) \in \mathcal{P}(G_\delta)} K_i K_{i_2} \cdots K_{i_n} \circledcirc^{\leq 0} \phi = \\ &= K_i \circ \Big( \bigcap_{(i_1,\ldots,i_n) \in I_m^{\bar{*}}} K_{i_1} K_{i_2} \cdots K_{i_n} \Big) \circ \Big( \bigcap_{(i_1,\ldots,i_n) \in (I_{m-1})^{\bar{*}}} K_{i_1} K_{i_2} \cdots K_{i_n} \Big) \circ \\ &\quad \circ \cdots \circ \Big( \bigcap_{(i_1,\ldots,i_n) \in I_1^{\bar{*}}} K_{i_1} K_{i_2} \cdots K_{i_n} \Big) \circledcirc^{\leq 0} \phi = \\ &= K_i C_{I_m} C_{I_{m-1}} \cdots C_{I_1} \circledcirc^{\leq 0} \phi = \\ &= C_{I_m} C_{I_{m-1}} \cdots C_{I_1} \circledcirc^{\leq 0} \phi. \end{aligned}$$

Thus, once more, Corollaries 5.2 and 5.3 reduce to the above results in this case as well.

The analogous results of Ben-Zvi and Moses for the rest of the response problems that they define (general ordered response [4], weakly-timed response [7] and tightly-timed response [7]) may be readily derived from our results in a similar manner — the details are left for the reader.

Having surveyed all the above response problems, one property, which is common to all of them (as well as to the rest of the response problems defined by Ben-Zvi and Moses) should be spelled out explicitly: they are all representable as special cases of timely-coordinated response, using $\delta$ s.t. for each $(i,j) \in I^{\bar{2}}$, either $\delta(i,j) = \infty$, or $\delta(j,i) = \infty$, or $\delta(i,j) = -\delta(j,i)$, i.e. the difference between the response times of $i$ and $j$ is bounded either from one side at most, or tightly (i.e. specified exactly). We note that the absence of this property in timely-coordinated response introduced a significant amount of complexity into our analysis, both technically and conceptually, and that without it, the machinery with which we analyzed timely-coordinated response could have been significantly simplified. Incidentally, for an analysis of timely-coordinated response that follows and extends the synchronous causality ("syncausality") approach of Ben-Zvi and Moses for analyzing response problems (and which makes this statement about the complexity introduced by an arbitrary $\delta$ function more concrete), the reader is referred to [16, Chapter 6].

### D.2 Common Knowledge and Variants

For the duration of this section, fix a system $R \subseteq \mathcal{R}$, an event $\psi \in \mathcal{F}_R$ and a set of agents $I \subseteq \mathbb{I}$. As noted above, while all previously-studied variants of common knowledge that are surveyed in Appendix B (and other previously-studied variants of common knowledge, such as continuous common knowledge [18]) are defined as fixed points of functions on $\mathcal{F}_R$, this is not the case with timely common knowledge, which we define as a fixed point of a function on $\mathcal{F}_R^I$. Intuitively, as noted above, this stems from the asymmetry of timely coordination with regard to the requirements posed on the various agents. Given this intuition, one may expect $\delta$-common knowledge to reduce, for constant $\delta$ (i.e. symmetric constraints), to a non-tuple fixed point in some way, and to coincide in some sense with the previously-studied variants of common knowledge surveyed above. To show this, we first note that $C_I^\delta \psi = (K_i(\psi \cap \bar{\xi}_i))_{i \in I}$, where $\bar{\xi}$ is the greatest fixed point of the function $\tilde{f}_\psi^\delta : \mathcal{F}_R^I \to \mathcal{F}_R^I$ given by

$$\tilde{f}_\psi^\delta : \quad (x_i)_{i \in I} \quad \mapsto \quad \left( \bigcap_{j \in I \setminus \{i\}} \circledcirc^{\leq \delta(i,j)} K_j(\psi \cap x_j) \right)_{i \in I}.$$

Next, we note that if indeed $\delta$ is a constant function attaining a nonnegative value, then it is straightforward to verify that $K_i(\psi \cap \xi_i) = K_i(\psi \cap (\cap \bar{\xi}))$ for every $i \in I$ (for $\bar{\xi}$ as defined above), yielding $C_I^\delta \psi = (K_i(\psi \cap (\cap \bar{\xi})))_{i \in I}$. Moreover, in this case $\cap \bar{\xi}$ is the greatest fixed point of $\cap \tilde{f}_\psi^\delta$. We now review the previously studied non-tuple variants of common knowledge surveyed above, and discuss when, and how, the above-described special case of $\delta$-common knowledge for constant $\delta$ generalizes them.

When $\delta \equiv \infty$, then by definition, $\delta$-coordination is equivalent to eventual coordination, $\cap \tilde{f}_\psi^\delta$ is the function presented in Theorem B.5(1), and thus $C_I^\delta \psi = (K_i(\psi \cap C_I^\lozenge \psi))_{i \in I}$. In addition, in this case Theorem 4.6(1,4,5) implies Theorem B.5.

Reducing our results for timely common knowledge to $\varepsilon$-common knowledge is somewhat more delicate. Assume, for the remainder of this section, that $\delta \equiv \varepsilon$ for some finite $\varepsilon \geq 0$. (Recall that for $\varepsilon = 0$, $\varepsilon$-coordination is equivalent to perfect coordination and Theorem B.6 reduces to Theorem B.4.)

In general, $\varepsilon$-coordination is a stricter condition than $\delta$-coordination.[20] However, for a (coordinate-wise) stable ensemble, as well as for an ensemble consisting at most of one point per agent per run, $\delta$-coordination is equivalent to $\varepsilon$-coordination — this follows from observing that given a $\delta$-coordinated ensemble, taking only the first point (or in a continuous-time model, the infimal point) of each agent in each run (and no points for runs in which the original ensemble contained no points for said agent) yields an $\varepsilon$-coordinated (and hence also $\delta$-coordinated) ensemble. If we restrict ourselves to stable $\psi$ and to protocols exhibiting perfect recall, then by Claim C.25, every coordinate of $C_I^\delta \psi$ is stable. Under these conditions, it may be verified that $K_i C_I^\varepsilon \psi$, for every $i \in I$, is stable as well.[21] In this case, by Lemma C.27, $C_I^\delta \psi$ is the greatest fixed point of $g_\psi^\delta$ and thus, $C_I^\delta \psi = (K_i(\psi \cap \xi))_{i \in I}$, where $\xi$ is the greatest fixed point of $\cap \tilde{g}_\psi^\delta$, where $\tilde{g}_\psi^\delta$ is defined analogously to $\tilde{f}_\psi^\delta$, but using $\circledcirc^{\delta(i,j)}$ in lieu of $\circledcirc^{\leq \delta(i,j)}$. Analogously to the proof of Lemma C.27, but in a less cumbersome way (as $\delta < \infty$), it may be shown that in this case $C_I^\varepsilon \psi$ is the greatest fixed point of $\cap \tilde{g}_\psi^\delta$ as well, and thus $C_I^\delta \psi = (K_i(\psi \cap C_I^\varepsilon \psi))_{i \in I}$,[22] and hence Theorem 4.6(1,4,5) reduces to Theorem B.6. In the absence of stability of $\psi$, or in the absence of perfect recall (at least of the "relevant events"), things stop working so well. Indeed, as noted above, in such cases $\delta$-coordination does not necessarily coincide with $\varepsilon$-coordination, and consequently, examples may be constructed in which $C_I^\delta \psi \neq (K_i(\psi \cap C_I^\varepsilon \psi))_{i \in I}$.

The above discussion raises an interesting question: why have we not defined $C_I^\delta \psi$ as $(K_i \xi_i)_{i \in I}$ instead of defining it as $(K_i(\psi \cap \xi_i))_{i \in I}$? (for $\bar{\xi}$ the greatest fixed point of $\tilde{f}_\psi^\delta$.) Indeed, the connection between such a definition and the previously-studied variants of common knowledge is much cleaner to describe [16, Chapter 10], and it yields results broadly similar to those presented in this paper [16, Chapters 7,8]. Nonetheless, much like $(K_i C_I^\varepsilon \psi)_{i \in I}$, and somewhat like $(K_i C_I^\lozenge \psi)_{i \in I}$, such a definition does not seem to naturally lend to a characterisation along the lines of "the greatest $\delta$-coordinated ensemble contained in $\psi$",[23] making it more cumbersome to use than the definition we presented in Section 4.

---

[20] This stems from two main "reasons":

1. $\delta$-coordination is defined using $\circledcirc^{\leq \delta(i,j)}$ rather than $\circledcirc^{[-\delta(j,i),\delta(i,j)]}$, which we define to mean "at some time no earlier than $-\delta(j,i)$ from now and no later than $\delta(i,j)$ from now". It may be readily verified that all the results in this paper hold for such a definition as well, as long as this replacement is performed in the definition of $f_\psi^\delta$ as well. The only difference is that Claim C.25, stating that $\delta$-common knowledge is stable, yields to different proof strategies in this case, e.g. showing that $(\circledcirc^{\leq 0}(C_I^\delta \psi)_i)_{i \in I} \leq f_\psi^\delta((\circledcirc^{\leq 0}(C_I^\delta \psi)_i)_{i \in I})$ and applying the induction rule for timely common knowledge.

2. Timely coordination is based on pairwise constraints. The results presented in this paper may be quite readily generalized to deal with arbitrary timing constraints of various natures, such as, e.g. for some $J \subseteq I$, "For every $i \in J$ and for every $(r,t) \in e_i$, there exists a time interval $T \subseteq \mathbb{T}$ of length at most $\delta_J$, s.t. $t \in T$ and s.t. there exist $(t_j)_{j \in J} \in T^J$ satisfying $(r, t_j) \in e_j$ for every $j \in J$". (Whatever the timing constraints are, the generalized definition of $f_{\psi_i}^\delta$ simply intersects on all constraints pertaining to $i$.) Under such a generalization, $\varepsilon$-coordination is equivalent to $\delta$-coordination, when setting $\delta_I \equiv \varepsilon$ in the above constraint example, and when providing no further constraints. Furthermore, in this case the generalization of $\tilde{f}_\psi^\delta$ satisfies that $\cap \tilde{f}_\psi^\delta$ is the function presented in Theorem B.6(1), and thus the appropriate generalization of Theorem 4.6(1,4,5) reduces to Theorem B.6.

It remains to be seen whether such generalizations as described in this footnote are of any real added value.

[21] This may be proved by showing that given stability of $\psi$ and perfect recall, it holds that $\circledcirc^{\leq 0} C_I^\varepsilon \psi \subseteq E_I^\varepsilon(\psi \cap \circledcirc^{\leq 0} C_I^\varepsilon \psi)$.

[22] Another way to derive this equality is by using [12, Exercise 11.17(d)], which shows that, for every $i \in I$, if $\psi$ is stable and given perfect recall, $K_i C_I^\varepsilon \psi = K_i(\cap_{n \in \mathbb{N}}(\circledcirc^\varepsilon E_I)^n \psi)$, and to apply (2). It should be noted, though, that the proof hinted to by [12, Exercise 11.17(d)] strongly relies on a discrete modeling of time, and breaks down in a continuous-time model, unlike the proof that we sketch above.

[23] While the ensemble defined by eventual common knowledge of an event of the form $\lozenge \psi$ is the greatest eventually-coordinated $I$-ensemble $\bar{e}$ satisfying $\cup \bar{e} \subseteq \lozenge \psi$ (the proof of this statement is left to the reader), we note that analogous characterisations for the ensembles defined by $\varepsilon$-common knowledge and by eventual common knowledge (of events not necessarily of the form $\lozenge \psi$) are, however, more elusive to phrase. (Moreover, parts 2–4 of Theorems B.5 and B.6 do not uniquely define these variants of common knowledge either.) In contrast, we note that $(K_i(\psi \cap C_I^\varepsilon \psi))_{i \in I}$ (resp. $(K_i(\psi \cap C_I^\lozenge \psi))_{i \in I}$) may be naturally characterised as the greatest $\varepsilon$-coordinated (resp. eventually-coordinated) $I$-ensemble whose union is contained in $\psi$.